\documentclass[aps,prl,twocolumn,superscriptaddress,groupedaddress]{revtex4}  
\usepackage{graphicx}  
\usepackage{dcolumn}   
\usepackage{bm}        
\usepackage{amssymb}   
\usepackage{amsmath}
\usepackage{bbm}
\usepackage{lipsum}
\usepackage{multirow}
\usepackage{array}
\newcolumntype{L}{>{\centering\arraybackslash}m{4.4cm}}
\newcolumntype{M}{>{\centering\arraybackslash}m{2.2cm}}
\newcolumntype{V}{>{\centering\arraybackslash}m{4cm}}
\usepackage{xcolor}
\usepackage{algorithmic}
\usepackage{algorithm}

\newtheorem{definition}{Definition}
\DeclareMathOperator*{\argmin}{arg\,min} 

\begin{document}

\widetext

\title{Noise-Robust Modes of the Retinal Population Code have the Geometry of ``Ridges" and Correspond with Neuronal Communities}
\author{%
Adrianna R. Loback\textsuperscript{1*},
Jason S. Prentice\textsuperscript{1},
Mark L. Ioffe\textsuperscript{2},
Michael J. Berry II\textsuperscript{1,3}
\\
\bigskip
\textbf{1} \emph{Princeton Neuroscience Institute, Princeton University, Princeton, NJ 08544, USA}
\\
\textbf{2} \emph{Department of Physics, Princeton University, Princeton, NJ 08544, USA}
\\
\textbf{3} \emph{Department of Molecular Biology, Princeton University, Princeton, NJ 08544, USA}
\\
\normalsize \href{mailto:}{*adrianna@princeton.edu} 
\bigskip
}
\date{\today}

\begin{abstract}
An appealing new principle for neural population codes is that correlations among neurons organize neural activity patterns into a discrete set of clusters, which can each be viewed as a noise-robust population \emph{codeword}.  
Previous studies assumed that these codewords corresponded geometrically with local peaks in the probability landscape of neural population responses.   
Here, we analyze multiple datasets of the responses of $\sim$$150$ retinal ganglion cells and show that local probability peaks are absent under broad, non-repeated stimulus ensembles, which are characteristic of natural behavior.  
However, we find that neural activity still forms noise-robust clusters in this regime, albeit clusters with a different geometry.  
We start by defining a \emph{soft local maximum}, which is a local probability maximum when constrained to a fixed spike count.  
Next, we show that soft local maxima are robustly present, and can moreover be linked across different spike count levels in the probability landscape to form a \emph{ridge}.  
We found that these ridges are comprised of combinations of spiking and silence in the neural population such that all of the spiking neurons are members of the same neuronal \emph{community}, a notion from network theory.  
We argue that a neuronal community shares many of the properties of Donald Hebb's classic cell assembly, and show that a simple, biologically plausible decoding algorithm can recognize the presence of a specific neuronal community.  
\end{abstract}

\maketitle

It is now clear from ample experimental and theoretical evidence that neural circuits throughout the brain  encode and transmit information using large populations of neurons \citep{Kalaska1992,Churchland1994,Sakurai1996,Nicolelis1997,Ghazanfar1997,Fujisawa2008,Puchalla2005,Truccolo2010,Harvey2012,Thesaurus2015}. 
Yet while the manner in which information is represented by single neurons has been intensively studied \citep{Perkel,Rieke1999,Dayan2001}, the empirical nature of neural population codes is still a topic of active investigation.  The advent of new experimental technologies that enable simultaneous recording from hundreds or even thousands of neurons \citep{Greenberg,Schlens2009,MarreSpikeSorting,Misha2013} has opened up exciting new possibilities to study this important question.  Fundamental to most conceptual approaches  is, as with the single neuron case, characterizing the probability distribution over all neuronal responses. The key additional issue for the multi-neuron scenario however is the nature of correlations among neurons, which fundamentally shapes the probability distribution of population activity.

So how do correlations affect the code of large neural populations? There are several ideas that have arisen from the past computational neuroscience literature. The oldest is that positive noise correlations can severely limit the encoded information, because they prevent large populations from averaging over the independent noise of neurons \citep{Zohary1994}.  However, this effect can be minimized if the noise correlations are orthogonal in the space of neural activity to the correlations induced by common stimulation \citep{Shamir2006, Moreno2014}. In fact, positive correlations with the right structure can even greatly enhance the encoded information \citep{Rava2014}.  
What these past studies have in common is that they have focused on how correlations quantitatively affect coding fidelity.  A second, yet less well-studied question, is how correlations affect the \emph{qualitative structure} of the neural population code.     

In the vein of this second approach, a recent idea is that correlations may organize neural population activity into a discrete set of clusters that constitute different ``codewords" \citep{Tkacik2010, Tkacik2014, Prentice2016, Kiani2007, Thesaurus2015}.  There are three notable advantages of having a code with this type of structure.  
Firstly, since the receptive field properties of multi-neuronal codewords have been shown to differ significantly from those of constituent single neurons \citep{Dan1998, Schnitzer2003, Schneidman2011, Prentice2016}, such a clustering operation constitutes a non-trivial computation that potentially changes the feature basis set used at each stage of a neural information processing pathway.  
Secondly, downstream neurons could use unsupervised learning mechanisms to identify these clusters (and hence learn the new feature basis set).  
Moreover, the identification of these clusters involves forming a map from many neural activity patterns onto the same cluster, or codeword.  Due to this many-onto-one mapping, these population codewords can exhibit \emph{error correction} - namely, that the codeword is more reliably activated by the stimulus than individual activity patterns \citep{Prentice2016}. 

Error correction is an idea that originates from communications engineering, which involves the design of codes for data transmission and storage that enable accurate decoding even when the transmitted information has been corrupted by noise \citep{Shannon1948, Roth2006}.  Traditionally, this robustness is achieved by introducing redundancy to partition the space of possible output patterns, so that all noise-corrupted versions of the same input message reliably map to the same subset. Consistent with this notion, the retinal code has been found empirically to be highly redundant \citep{Puchalla2005}.  Furthermore, sensory maps in the cortex, such as V1, have many more neurons than do their subcortical sources, implying that these brain areas are even more redundant \citep{Barlow2001}. Error correction is an appealing principle, because it offers a way to bridge the gap between the noisy activity of neurons and the fact that our perception can be quite deterministic \citep{Rava2014}.

Due to the combinatorial explosion of possible neural population responses, it is generally intractable to determine the empirical probability distribution of joint activity. 
This fundamental limitation necessitates a modeling approach. 
Within computational neuroscience, there are two distinct methodologies that have been used to create probabilistic models of network activity.  The traditional approach - which includes generalized linear models (GLMs) - attempts to explicitly capture the dependence of neuronal responses on the stimulus \citep{Pillow2008}. 
These types of models have been referred to as ``encoding models" \citep{Vidne2012}.  
The second approach, which we call the ``activity model" approach \citep{Prentice2016}, involves directly modeling the structure of population activity, without any reference to the stimulus.  
An important advantage of activity models is that they correspond to the unsupervised problem actually faced by downstream brain areas, which lack direct access to the external stimuli.  

One popular class of activity models comes from the Maximum Entropy (MaxEnt) principle (see Appendix A.1) \citep{Schneidman2006, Ohiorhenuan2010, Vasquez2011, Ganmor2011, Nasser2013, Watanabe2013, Mora2015}. 
Previous work which fit the K-Pairwise MaxEnt model to activity measured from retinal ganglion cells reported a proliferation of basins in the associated energy landscape \citep{Tkacik2014}. These basins correspond to local peaks in the joint response probability landscape (see Fig 1), and can be found by an iterative algorithm that changes the firing state of a single neuron to increase the probability at each step (Fig. 1C).  
It was suggested that these local probability peaks could be a candidate for neural population codewords \citep{Tkacik2014}. 

\begin{figure}[ht!]
\centering
\includegraphics[scale=0.54]{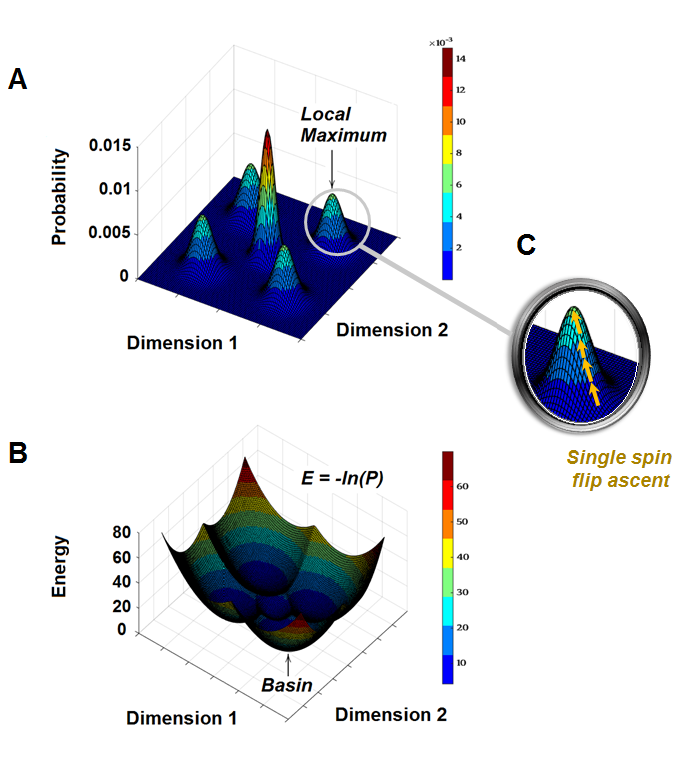}
\caption{ Schematic illustrating the concept of local maxima and basins.
(A) Cartoon illustrating local maxima in a probability landscape. For ease of visualization, the full space of population activity patterns is depicted as a continuous, $2D$ space. The actual domain of the joint probability mass function considered here is discrete and high-dimensional (in particular, an $N$-dimensional hypercube).
(B) Cartoon of basins in the energy landscape corresponding to the probability landscape shown in panel ($A$). A response state $\vec{\sigma}$ that is a local probability maximum equivalently corresponds to a basin in the associated energy landscape, since the energy $E(\vec{\sigma}) \equiv -\ln P(\vec{\sigma})$. 
(C) Local maxima in the probability landscape are found via an iterative single spin flip ascent algorithm. Each yellow arrow denotes one iteration.}
\label{fig1}
\end{figure} 

In this study, we started by uncovering an important problem.  Previous studies that have identified putative population codewords as corresponding to local peaks in the probability landscape used experimental data in which a short stimulus segment was repeated many times.   
However, when we applied the same analysis methods to retinal ganglion cell data under stimulation by non- and mildly-repeated stimulus ensembles, we found almost no local probability peaks.  
On the other hand, a recent study that used a hidden Markov model (HMM) to describe the ganglion cell population's probability landscape was able to find clusters in the population activity space - called ``collective modes" - under non-repeated visual stimulation \citep{Prentice2016}. 
Importantly, these collective modes were shown to provide a new feature basis set (including some modes with orientation selectivity), and to exhibit error correction, supporting their candidacy as neural population codewords.   
Building on this previous work, we confirmed that putative population codewords are identifiably present under non- and mildly-repeated visual stimulation.  
Our first result is that they do not correspond with local peaks in the probability landscape.   

Motivated by this result, we next sought to investigate the geometry of population codewords for the non-repeated stimulus regime. 
To do so, we introduced a new notion of structure, here termed \emph{soft local maxima}, which are local maxima in the space of all neural activity patterns that are restricted to have the same spike count.  These soft local maxima were robustly present.  
Using a novel numerical approach, we then explored their organization across spike count levels. This approach led to an algorithm that links together different soft local maxima into a discrete \emph{ridge}, which is a geometric feature of the joint probability landscape of neural population activity. We found that there was a close correspondence between these ridges and the population codewords found by the hidden Markov model.  
We argue that depending on the behavioral context, experiments designed with either repeated or non-repeated stimulus ensembles can be relevant, and therefore both geometric structures - local peaks and ridges - can be important for the population code. 
Finally, we realized that these ridges correspond with \emph{neuronal communities}, a notion from network theory \citep{Fortunato2010}.  Our results thus suggest a unified picture linking statistically-defined ``codewords" of the retinal population code, geometric structure in the joint response probability landscape, and community structure within the neuronal population. 

\section{Methods} 

\subsection{Experimental Procedures}
\subsubsection{Electrophysiology}
We analyzed recordings from larval tiger salamander (\emph{Ambystoma tigrinum}) retinal ganglion cells responding to either naturalistic or white noise checkerboard movie stimulus ensembles. Raw voltage traces were digitized and stored for offline analysis using a 252-channel preamplifier (MultiChannel Systems, Germany). Offline spike sorting was performed using custom software. Detailed recording and spike-sorting  methods are described in \cite{MarreSpikeSorting}.  
For all experiments, only the responses of neurons that passed the standard tests for waveform stability and lack of refractory period violations were included. 

\subsubsection{Visual Stimulus Display}
Stimuli were projected onto the array from a CRT monitor at a frame rate of 60 Hz, and gamma corrected for the display.  
We presented two stimulus classes, in a total of four different experiments: a natural movie (three experiments), and a binary white noise checkerboard (one experiment).  
The natural movie consisted of a 7-min gray scale recording of leaves and branches blowing in the wind. We conducted the natural movie experiments with three different designs: (1) in the first, which we refer to throughout ``Movie $\#1$", the full 7-min movie was looped (i.e. sequentially repeated) ten times, for a total movie length of approximately $70$ min; (2) in the second, which we refer to as ``Movie $\#2$", the movie was looped five times, but with a different pixel centered on the recording area for each repeat; (3) in the third, which we refer to as ``Movie $\#4$", we interleaved 73 unique movie segments, each 60 s in duration, with 73 repeats of a fixed 60-s ``target" movie segment.  
Design (2) was constructed this way to provide a non-repeated stimulus: since the patch of retina recorded by the multi-electrode array subtended only a small portion of the stimulus, the retinal input was effectively non-repeated over the full recording session.  
The binary white noise stimulus, which we refer to throughout as ``Movie $\#3$", consisted of a $40 \times 40$ array of $58 \: \mu$m squares with light intensity randomly selected to be either bright or dark every 30 ms. 
This checkerborad stimulus was formatted into 30 s periods that alternated between non-repeated (unique) stimulation and a repeated 30-s ``target" movie segment.  
A total of 69 unique movie segments and 69 repeats of the same target segment were presented. 

\subsection{Data Preparation}
For each dataset, we discretized spike trains into 20 ms time bins, as in previous work \citep{Thesaurus2015,Schneidman2006,Tkacik2014,Prentice2016}. This produces a sequence of binary population response vectors, which we denote as $\vec{\sigma}(t) \equiv \left( \sigma_1(t),\cdots,\sigma_i(t),\cdots \sigma_N(t)\right) \in \{0,1\}^N$, where $i = 1,\cdots,N$ labels the neuron identity and $t$ the time bin. We set $\sigma_i(t) = 1$ if neuron $i$ fired at least one spike within time bin $t$, and $0$ otherwise. 

\subsection{Fitting the Tree Hidden Markov Model to Data}
In this study, we fit the statistical model recently introduced in \cite{Prentice2016}, called the \emph{Tree hidden Markov model} (HMM), to each of our datasets.  The Tree HMM models the temporal sequence of observed neural population responses as:
\begin{eqnarray} \label{eq:3}
P \Big( \vec{\sigma}(1),\cdots,\vec{\sigma}(T) \Big) = \sum_{\vec{\alpha} \in [m]^T} \left(
\prod_{t=1}^T Q_{\alpha_t} \Big( \vec{\sigma}(t) \Big) \Gamma_{\alpha_{t-1},\alpha_t}
\right)
\end{eqnarray}

\noindent where $T$ denotes the total number of time bins, $\vec{\alpha} \equiv (\alpha_1,\cdots,\alpha_t,\cdots,\alpha_T)$ denotes the sequence of latent states (codewords), and $Q_{\alpha}(\cdot)$ denotes the emission distribution for mode $\alpha \in [m] \equiv \{1,\cdots,m\}$.  In practice, we assume that the transition matrix $\Gamma$, which has entries $\Gamma_{\alpha_{t-1},\alpha_t} \equiv P(\alpha_t | \alpha_{t-1})$, is stationary (i.e. time-independent). 
Each emission distribution $Q_{\alpha}$ in the model is a tree graphical model \citep{Meila2006}.  
See Appendix A.2 and \citep{Prentice2016} for further details. 

Note that when exploring the probability landscape modeled by the Tree HMM, we used the stationary distribution $\vec{\psi}$ of the fit Markov chain.  
Formally, $\vec{\psi}$ is defined as the left eigenvector of the transition matrix $\Gamma$ with unity eigenvalue (i.e. $\vec{\psi}^{\intercal} \Gamma = \vec{\psi}^{\intercal}$) and satisfies $\vec{\psi}^{\intercal} \vec{1} = 1$.  
This allows us to write the modeled joint probability, $P_{\text{stationary}}$, for a single population response $\vec{\sigma} \in \{0,1\}^N$ as:
\begin{eqnarray} \label{eq:7}
P_{\text{stationary}} \left( \vec{\sigma} \right) = \sum_{\alpha=1}^m \psi_{\alpha} Q_{\alpha} (\vec{\sigma})
\end{eqnarray}

\subsection{Finding Local Maxima}
In the present work, we use the same definition of local maxima of the joint response probability landscape that was used previously in \cite{Tkacik2014}. 
That is, we defined local probability maxima as the single-flip-stable ascent patterns.
This definition is equivalent to what is termed a ``$(\delta$=$1,\rho$=$\text{Hamming distance})$-mode" in applied math \citep{ChenModes}:  
\begin{definition}
A point is a $(\delta, \rho)$\emph{-mode} if and only if its probability is higher than all points within distance $\delta$ under a distance metric $\rho$.
\end{definition}

\noindent To find local maxima of the high-dimensional joint response probability landscape, we used the same iterative algorithm as in \cite{Tkacik2014}.  We refer to this algorithm as ``single spin flip ascent" (see Appendix A.5.1 for details). 

\section{\label{sec:level1}Results}

\subsection{Population Codewords do not Correspond to Local Maxima for the Low-Repeat Stimulus Regime}
Previous studies that used the Maximum Entropy framework (see Appendix A.1) to explore the probability landscape of neural population activity have reported a proliferation of local probability maxima \citep{Tkacik2014}.  
However, whether or how this feature depends on the stimulus ensemble was unknown. 
One potential confound of these studies, is that they have exclusively used highly-repeated stimulus ensembles.   
For a highly-repeated stimulus ensemble, a small set of points in stimulus space is sampled on many trials.  
It is thus plausible that the previously reported local maxima correspond to `average' response patterns elicited by the stimuli comprising the short movie segment that was repeated, with the width of the local peaks being attributable to scatter around that average due to neural noise (see section 4 for details).

To investigate whether a proliferation of local maxima is a general feature that is independent of the stimulus repeat structure, we first applied the same K-Pairwise MaxEnt model used in \cite{Tkacik2014} to the measured responses of $N=128$ retinal ganglion cells to Movie $\#1$, which was a mildly-repeated natural movie stimulus. 
In response to Movie $\#1$, the mean firing rate averaged over ganglion cells, $\langle r(t) \rangle$, was $2.70 \pm 0.21$ spikes/s/neuron (mean $\pm$ SEM).  
The probability of a cell firing a spike in any given time bin $t$, which we denote $p(t) \equiv \langle r(t) \rangle \triangle t$ (where $\triangle t = 0.02$ s), was $0.054$. 
The measured responses were thus within the sparse firing regime. 

After fitting the MaxEnt model to the data (see Appendix A.1), we then searched for local maxima of the modeled probability landscape using the same numerical method as in \citep{Tkacik2014}, which we term ``single spin flip ascent".  
In brief, this involves taking each population activity pattern observed in the data, and moving ``uphill" on the modeled joint response probability landscape. 
The termination patterns of this algorithm are by definition local maxima of the probability landscape, when only allowing the state of a single neuron to change in each iteration.  
Note that this choice of defining a local maximum as the single-flip-stable ascent pattern establishes an upper bound on the number of local maxima \citep{ChenModes}. 
The single spin flip ascent algorithm can be thought of as a mapping from the $N$-dimensional population response space to the set of local probability maxima.

\subsubsection{Scarcity of Local Maxima for Low-Repeat Stimulus Ensembles}
We applied the single spin flip ascent algorithm to the probability landscape obtained by fitting the MaxEnt model to the Movie $\#1$ dataset.  This resulted in 65 unique local probability maxima (Fig 2A). 
However, of the 175,002 observed population responses in the data, $99.71\%$ were mapped via single spin flip ascent to the all-silent local maximum (i.e. $\vec{0}$), and $0.037\%$ were mapped to the rank-2 local maximum. 
This is in stark contrast to the $\sim$$50\%$ of responses that were mapped via single spin flip ascent to the all-silent local maximum for the highly-repeated stimulus dataset in \cite{Tkacik2014}. 

\begin{figure}[!h]
\centering
\includegraphics[scale=0.45]{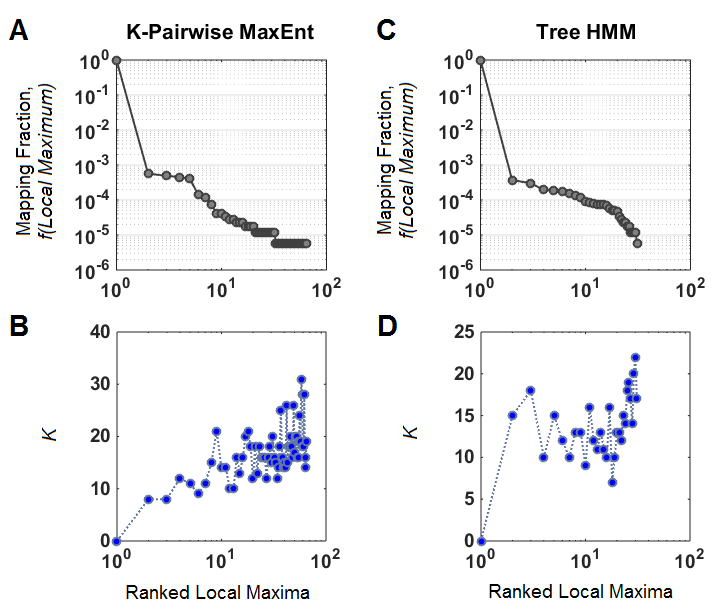}
\caption{Local maxima results obtained for the dataset of $128$ ganglion cells responding to Movie $\#1$, using either the K-Pairwise Maximum Entropy model (8385 parameters) (A,B) or Tree hidden Markov model (8379 parameters) (C,D) as the underlying probability model. 
(A) Log-log plot of the proportion of the 175,002 population responses observed in the data that were mapped via single spin flip ascent - denoted as the ``mapping fraction", $f(\text{Local Maximum})$ - to the corresponding unique local maximum indicated on the $x$-axis (ranked). 
(B) The spike count, denoted by $K \equiv \sum_{i=1}^N \sigma_i$, for each of the 65 unique local maxima identified ($x$-axis, ranked).
(C,D) The panels are as described in (A) and (B), but for the hidden Markov model results.
}
\label{fig2}
\end{figure}

We also fit the Tree hidden Markov model to this dataset (see Appendix A.2), and likewise performed single spin flip ascent on the corresponding probability landscape. 
Since the number of free parameters constitutes an important consideration for model comparison, we selected the number of HMM latent states so as to match the total number of free parameters as closely as possible to the K-Pairwise MaxEnt model.  
As shown in Fig 2C, $31$ unique local maxima were found after performing single spin flip ascent on the probability landscape modeled by the HMM. Of the 175,002 observed population responses in the data, $99.74\%$ were mapped to the all-silent local maximum, and $0.056\%$ were mapped to the rank-2 local maximum. 
This result is consistent with that obtained for the K-Pairwise MaxEnt model, indicating that the paucity of local maxima is an intrinsic feature of the probability landscape rather than an artifact of one specific model.

To investigate the limit of the non-repeated stimulus regime, we next performed the above procedure on a dataset of the responses of $152$ ganglion cells to Movie $\#2$, which was a \emph{non-repeated} natural movie stimulus.  
The mean firing rate elicited by Movie $\#2$, averaged over ganglion cells, was $1.10 \pm 0.07$ spikes/s/neuron (mean $\pm$ SEM).  
Correspondingly, the probability of a neuron firing a spike in any given time bin was again sparse, $p(t) = 0.022$.   
For this non-repeated stimulus dataset, performing single spin flip ascent on the probability landscape modeled by the Tree HMM identified only two local maxima (see Fig \ref{fig:S2} in Appendix D).   
Moreover, a staggering $99.95\%$ of the 90,001 observed population responses were mapped via single spin flip ascent to the all-silent local maximum. 

\subsubsection{A Systematic Analysis}

To probe the relationship between stimulus repeat structure and the prevalence of non-silent local maxima more systematically, we next performed an analysis that allowed us to take the number of stimulus repeats as a parameter. 
Specifically, two separate experiments were performed in which we presented one of two different movie stimuli: a binary white-noise checkerboard movie (Movie $\#3$), or a natural movie (Movie $\#4$).  Both movie stimuli were similarly designed to have unique movie segments that were \emph{interleaved} with repeated presentations of a ``target" movie segment (see section 2 for details).   
A schematic illustrating the experimental setup for the interleaved white-noise stimulus ensemble is shown in Fig 3A. 

We verified that for both datasets, each ganglion cell's average firing rate during the repeated vs. unique movie segments was statistically identical (see Fig \ref{fig:S1} in Appendix D). 
For Movie $\#3$, the average firing rate across ganglion cells recorded during the repeated segments and unique segments was, respectively, $0.86 \pm 0.08$ spikes/s/neuron and $0.84 \pm 0.08$ spikes/s/neuron (mean $\pm$ SEM).
Correspondingly, the probability of a neuron firing a spike in any given time bin was $p(t) = 0.0172$ and $p(t) = 0.0168$, respectively. 
For Movie $\#4$, the average firing rate across ganglion cells recorded during the repeated clips and unique clips was, respectively, $1.40 \pm 0.01$ spikes/s/neuron and $1.46 \pm 0.01$ spikes/s/neuron.  
The probability of a neuron firing a spike in any given time bin was $p(t) = 0.028$ and $p(t) = 0.029$, respectively.

\begin{figure*}[!ht]
\includegraphics[scale=0.28]{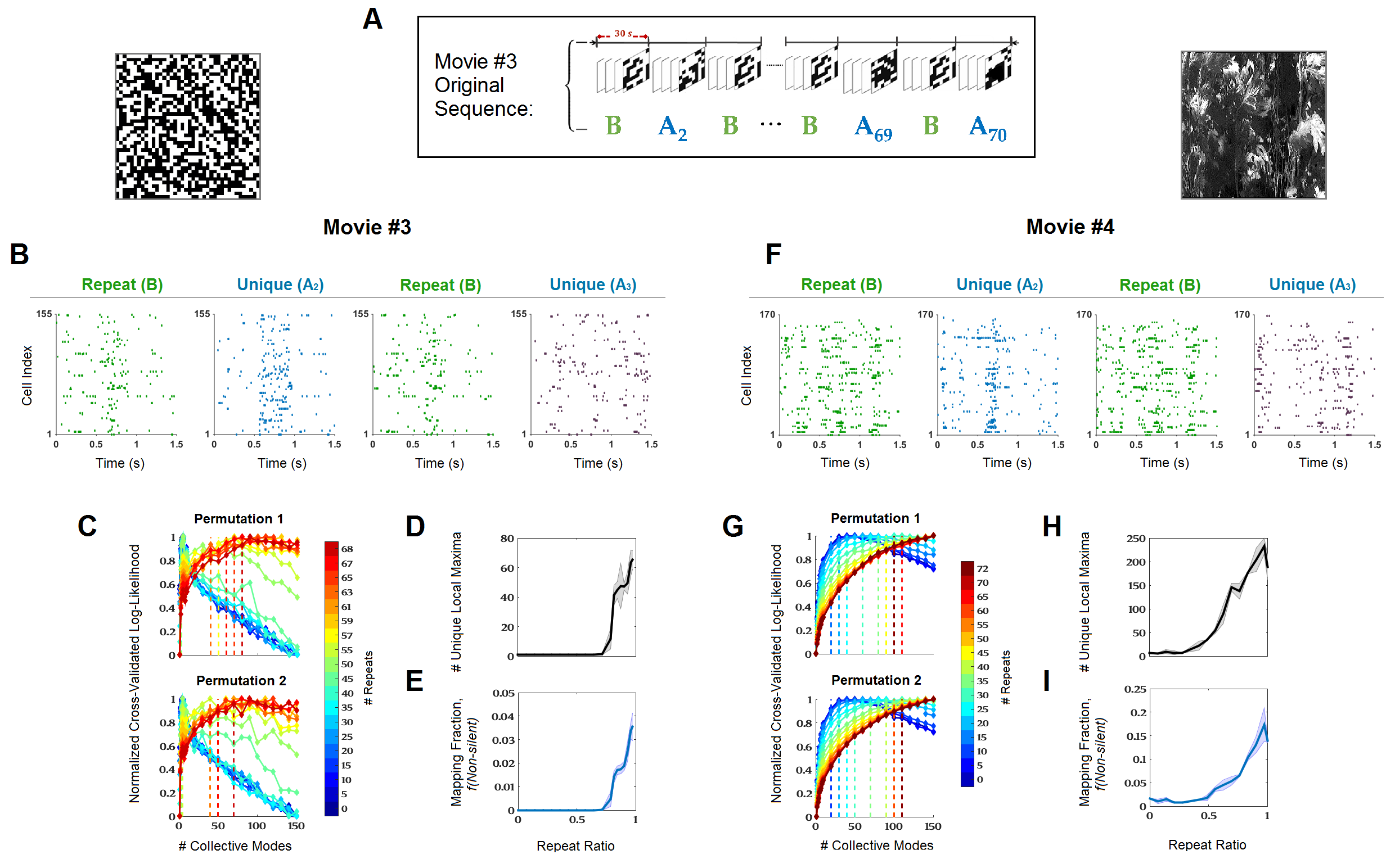}                           
\caption{Experimental design and results for the parametric repeat analysis.  
(A) Cartoon of the movie stimulus design.  Unique movie segments (denoted by blue $A$'s) were interleaved with repeated pesentations of a ``target" movie segment (denoted by green $B$'s).  
(B-E) Results for the dataset of $155$ ganglion cells responding to Movie $\#3$. 
(B) Examples of raw spike rasters elicited by repeated (green) versus non-repeated movie segments (blue and purple).  
(C) Normalized cross-validated log-likelihood (CV-LL, $y$-axis) vs. the number of Tree HMM latent states for each repeat ratio (see color key). 
Dashed lines denote the optimal number of latent states (see \emph{Methods}). 
(D) Total number of unique local maxima (black) found vs. repeat ratio. 
Error bars denote SEM. 
(E) Proportion of the 105,000 population responses observed during each respective subset movie that were mapped via single spin flip ascent to a non-silent local maximum. 
(F-I) Results for the dataset of $170$ ganglion cells responding to Movie $\#4$.
}
\label{fig3}
\end{figure*} 

For each of the two movie stimuli, we then generated ``subset movies", which were comprised of a subset of the movie segments that appeared in the original full-length movie (see Appendix A). 
Different subset movies included a different ratio of the number of repeated target segments to unique movie segments.
Note that the duration of each subset movie was the same, so as to eliminate the potentially confounding effect of different stimulus durations on our sampling of population neural activity. 
To ensure that there was no dependence on the specific choice or ordering of the movie segments included, we also performed multiple independent random permutations to generate the subset movie corresponding to each repeat ratio.  
For each permutation and for each repeat ratio, we then fit the Tree hidden Markov model to the set of ganglion cell population responses observed during the corresponding subset movie. 
The cross-validated log-likelihood results (used to select the optimal number of latent states, or ``collective modes"; see Appendix A.2) for different repeat ratios are shown in Fig 3C and 3G.  For both datasets, there was a general shift toward a larger optimal number of collective modes as the number of included repeats increased.

After fitting the Tree hidden Markov model to the sequence of ganglion cell population responses observed for each subset movie, we then carried out the single spin flip ascent algorithm on the modeled probability landscape.  
For both datasets, we observed a predominantly monotonic increase in the number of unique local maxima as a function of repeat ratio.  Specifically, for the white-noise (Movie $\#3$) dataset, we found $1 \pm 0$ unique local maximum corresponding with the case of 0 included repeats, which increased to $65 \pm 6$ local maxima when all movie segments were repeats (Fig 3D). 
Likewise, for the interleaved natural movie (Movie $\#4$) dataset, increasing the repeat ratio from 0 to 1 corresponded with a drastic increase in the average number of identified local maxima, from $6 \pm 0$ to $232.5 \pm 16.5$ (Fig 3H).   

Moreover, for both datasets, the weight of non-silent local maxima - that is, the percentage of observed population responses that were mapped by single spin flip ascent to a non-silent local maximum - increased mostly monotonically as a function of the repeat ratio.  
In particular, for the interleaved white-noise dataset we found that the weight of non-silent local maxima increased from $0\%$ for the non-repeated case, to $3.54\% \pm 0.59\%$ for the all-repeat case (Fig 3E).  For the interleaved natural movie dataset, the effect was stronger, with the corresponding value being $1.63\% \pm 0.01\%$ for the non-repeated case, and $17.34\% \pm 3.54\%$ for the all-repeat case (Fig 3I). 

In summary, we have found that the qualitative structure of the probability landscape of ganglion cell population activity strongly depended on the stimulus ensemble.  
Specifically, within the non- and mildly-repeated stimulus regimes, we found that for both artificial and natural movie stimuli the modeled response probability landscape was essentially comprised of a single global peak, arranged around the all-silent activity pattern, $\vec{0}$.  
This result refutes the idea that local maxima of the probability landscape could be a candidate for noise-robust population codewords for the \emph{low-repeat} stimulus regime.  
This is because using local peaks as the codewords for this regime would result in mapping the entire population response space to the all-silent codeword, corresponding with an encoding scheme with a prohibitively low channel capacity. 
However, when particular stimuli were repeated sufficiently, this resulted in the incorporation of non-silent local maxima in the probability landscape. 

\subsection{Probing the Geometry of Collective Modes}
The results in section 3.1 demonstrate that local maxima are an improbable candidate for population codewords in the low-repeat stimulus regime, which is a stimulus regime that is characteristic of natural behavior.   
Recent work has suggested that there may be a better codeword candidate.  
In particular, by applying the Tree hidden Markov model, it was found that the $N$-dimensional response space of ganglion cell population activity is organized into clusters, called ``collective modes" \citep{Prentice2016}.   
These collective modes were shown to provide a new feature basis set different from the receptive fields of constituent ganglion cells, and were moreover shown to exhibit error correction.    
Importantly, this error correction property was demonstrated for the \emph{non-repeated} stimulus regime.  

The hidden Markov model approach introduced in \cite{Prentice2016} is a statistical modeling framework that formalizes the fact that the population codewords are not directly observable by representing them as latent (hidden) states, which have a one-to-one correspondence with the collective modes.  
Since this approach is statistical, it importantly does not make explicit assumptions about the geometry of the population codewords.  
This thus leaves open the question: What is the geometric correlate of the collective modes in the response probability landscape?  
As fitting the hidden Markov model is not a biologically plausible computation, it would be useful to know whether the statistically-defined collective modes correspond to other structural correlates that could potentially be detected by biologically plausible algorithms.  

We have seen in section 3.1 that non-silent local maxima are exceedingly rare for the low-repeat stimulus regime (Figs 2 and 3).  
Yet many collective modes are identified in both the low- and high-repeat stimulus regimes (Fig 3) \citep{Prentice2016}.  
Together, these findings indicate that, for the low-repeat stimulus regime, collective modes do \emph{not} have the geometry of local peaks in the probability landscape. 
Motivated by this result, we next sought to characterize the geometry of the collective modes.   

\subsubsection{The Role of Sparseness} 
As a first step, we note that neural activity of retinal ganglion cells was very sparse across all of the stimulus ensembles tested, both artificial and natural.  
This leads to a simple intuition about why we see essentially no local peaks in the probability landscape in the low-repeat stimulus regime.  
Due to the sparseness of neural activity, it may be quite unlikely that an activity pattern with $K$ spikes has higher probability than a pattern with $K-1$ spikes formed by switching one of its spiking neurons to silent.  

To gain a sense of how strongly sparseness affects the probability of activity patterns, we plotted the empirical probability of finding activity patterns with $K$ spikes (Fig 4).  
For all stimulus ensembles examined, we found that this probability decreased monotonically with increasing $K$.  
Moreover, when we fit a log-linear model to this function at low $K$ values ($0 \leq K \leq 4$), the slope for the moderately-repeated Movie $\#1$ dataset was a factor of $\sim$$50$-fold decrease in probability per extra spike.  
For the non-repeated Movie $\#2$ and non-repeated white-noise checkerboard dataset, the fit slope was even higher, being a factor of an $80$-fold decrease and a $90$-fold decrease in probability per extra spike, respectively.  
This indicates that sparseness controls the probability of activity patterns quite powerfully. 

\begin{figure}[!h]
\includegraphics[scale=0.45]{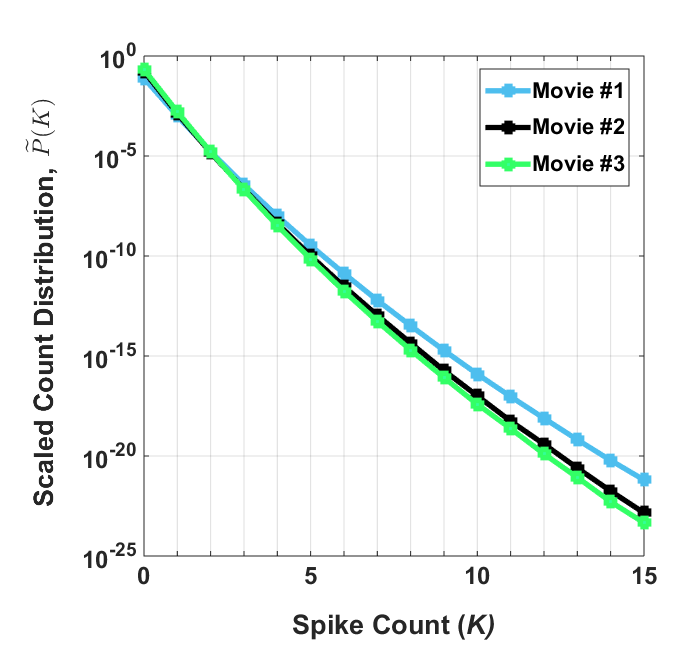}
\centering
\caption{The empirical ``scaled count distribution", $\widetilde{P}(K)$, as a function of spike count $K$. 
The scaled count distribution is defined as the empirical probability of observing a population response with $K$ spikes, normalized by the analytical number of possible joint response patterns with spike count $K$ (see Appendix A). 
Results are shown for three of the different datasets analyzed in the present paper: Movie $\#1$ (blue; $N=128$ ganglion cells), Movie $\#2$ (black; $N=152$ ganglion cells), and the non-repeated version of Movie $\#3$ (green; $N=155$ ganglion cells).  Note that the $y$-axis is shown on a log scale. 
}
\label{fig4}
\end{figure}

The powerful impact of sparseness led us to postulate that while local maxima may be largely absent in the full probability landscape, perhaps there are robust local maxima within the restricted space of all activity patterns with shared spike count, $K$.  
More generally, we can restate this postulate as one in which we assume that there is low-dimensional structure in the high-dimensional probability landscape that could be exploited by downstream brain areas.  
After all, there are various ``no free lunch" theorems that show that, in the absence of low-dimensional structure, very little can be done in high-dimensional learning problems with limited samples \citep{NRWY12}. 

\subsection{``Soft" Local Maxima} 
To explore this idea, we investigated whether the joint response probability landscape for the non-repeated stimulus regime contains what we term \emph{$K$-soft local maxima}.  
Intuitively, a $K$-soft local maximum is a local probability maximum when we restrict our search to the metric subspace (of the full joint response space, which is the $N$-dimensional Hamming cube) defined on the set of all activity patterns with fixed spike count, $K$.  We will subsequently refer to each such metric subspace as the \emph{$K$-th spike count level}.  

The geometric intuition behind this definition is illustrated in Fig 5.  
Specifically, in its most simplified form, our motivating intuition was that the global probability landscape resembles a ``mountain" with the global peak given by the all-silent state.  Coming down the mountain (in the direction of higher $K$) are a number of ridges.  
Along each ridgeline, the probability is a local maximum of activity patterns at constant spike count, $K$ (blue and green curves, Fig 5B).  
But the strong decrease of probability with $K$ (Fig 5C) prevents these activity patterns (such as points $\gamma_{10}$ and $\gamma_{15}$, Fig 5A) from being true local maxima.  
We will show in subsequent sections that this is a qualitatively viable picture of the actual, high-dimensional joint response probability landscape in the low-repeat stimulus regime.  

\begin{figure}[!h]
\includegraphics[scale=0.4]{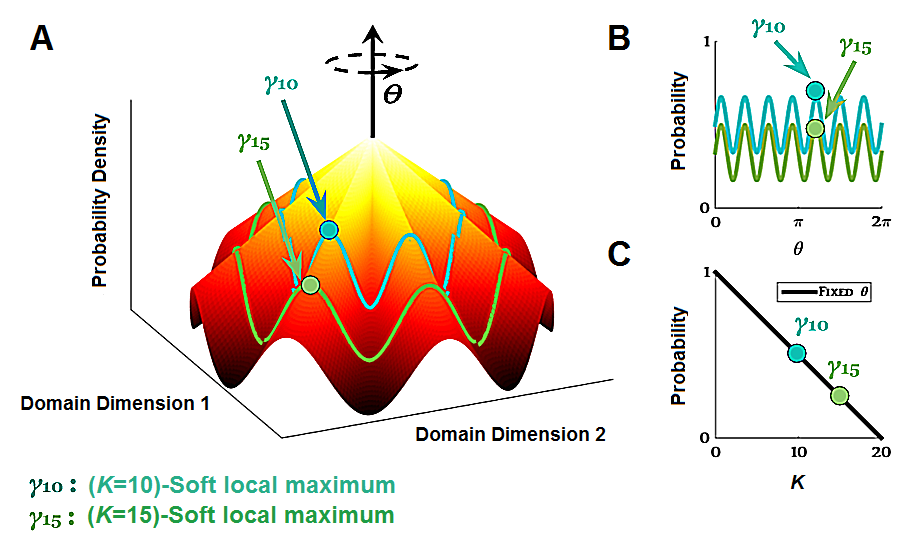}
\centering
\caption{Schematic illustrating the concept of $K$-soft local maxima and ridge points.
(A) Cartoon of a multivariate probability mass function and its ``ambient" (i.e. prior to projection) domain. 
For ease of visualization, the ambient domain is shown in 2D, and is represented in polar coordinates. Here we denote the radial coordinate by $K$, and the polar coordinate by $\theta$. 
The analogue of the $K$-soft local maxima defined in Def. (\ref{def:1}) for this conceptual example are the points that are local maxima of the conditional probability $P(\theta|K)$, i.e. after we project onto a fixed value of $K$. 
For illustration, two example $K$-soft local maxima are shown, denoted by circles and labeled $\gamma_{10}$ (blue) and $\gamma_{15}$ (green; see arrows). 
(B) After projecting onto the radial coordinate for the respective values of $K$, we see that $\gamma_{10}$ is a local maximum of $P(\theta | K=10)$ and likewise that $\gamma_{15}$ is a local maximum of $P(\theta | K=15)$. 
(C) Illustrated is the projection onto the polar coordinate, for an arbitrary fixed value of $\theta$. 
Since soft local maxima are local maxima of a function defined on a lower-dimensional subspace of the original domain, they are candidates for a type of ``ridge point," and as shown in Panel (A) could theoretically form discrete ``ridges" that span across different $K$ levels. 
}
\label{fig5}
\end{figure}

If we wish to find nearby activity patterns that preserve the same spike count, then if one neuron's silence is `flipped' (changed) to spiking, this necessitates changing another neuron's spiking to silence (and vice versa). 
Thus, given an activity pattern with spike count $K$, its ``neighbors" in the $K$-th spike count level will be those that differ by a Hamming distance of 2.    
In other words, soft local maxima are the stable-probability-ascent patterns when we allow for flipping exactly two opposing neuron response states.  
Formally, this can be formulated as follows:

\begin{definition} \label{def:1}
A response state $\vec{\gamma} \in \{0,1\}^N$ is a \emph{$K$-soft local maximum} if and only if $P[\vec{\gamma}] > P[\vec{\sigma}]$ $\: \forall \: \vec{\sigma}$ such that $w_H(\vec{\sigma}) = w_H(\vec{\gamma})=K$ and $d_H(\vec{\sigma},\vec{\gamma}) = 2$, where $w_H(\vec{\sigma}) \equiv \sum_{i=1}^N \sigma_i$ denotes spike count (Hamming weight) and $d_H(\vec{\sigma},\vec{\gamma}) \equiv \sum_{i=1}^N | \sigma_i - \gamma_i |$ denotes Hamming distance.  
\end{definition}

\noindent Note that, to our knowledge, this concept is new in the neuroscience literature.  

In practice, we used an iterative algorithm that we term \emph{opposite sign neuron pair relaxation} to find the $K$-soft local maxima for a given spike count $K$ (see Appendix A.5.2 for details).  
Note that our choice of this algorithm is \emph{not} simply the next obvious ``higher order" search technique from the single spin slip ascent algorithm. 
Rather, it was motivated by the sparseness of neural activity, and follows directly out of necessity when imposing the constraint to preserve spike count.   

To check for the presence (or absence) of soft local maxima in the probability landscape, we implemented the opposite sign neuron pair relaxation algorithm for two large datasets. Based on the results of another analysis (see Appendix A.8), which demonstrated that the Tree HMM more accurately captures the empirical $(K=2)$-soft local maxima than the K-Pairwise MaxEnt model, we chose the Tree HMM as our underlying probability model.  
 
We first examined the dataset of $152$ ganglion cells responding to the non-repeated natural movie stimulus (Movie $\#2$).  
For all spike counts $K>1$ examined, we found multiple soft local maxima (Fig 6A, grey curve).  
Moreover, the number of identified unique soft local maxima monotonically increased with spike count.  
Since the opposite sign neuron pair relaxation algorithm is stochastic, we checked how robust the mapped set of unique $K$-soft local maxima was.  To do so, for each value of $K$ we performed $100$ independent iterations of the opposite sign neuron pair relaxation algorithm (see Appendix A.6).    
We then computed the mean pairwise overlap ratio between the identified sets of unique $K$-soft local maxima for each of the $\binom{100}{2}$ iteration pairs. 
As seen in Fig 6A (upper left-hand inset), for $1 \leq K \leq 3$, the identity of the mapped set of unique soft local maxima was perfectly conserved.  
For $4 \leq K \leq 7$, 
the mean pairwise overlap ratio was also large ($>94\%$), indicating a high degree of robustness.

\begin{figure*}[!ht]
\includegraphics[scale=0.37]{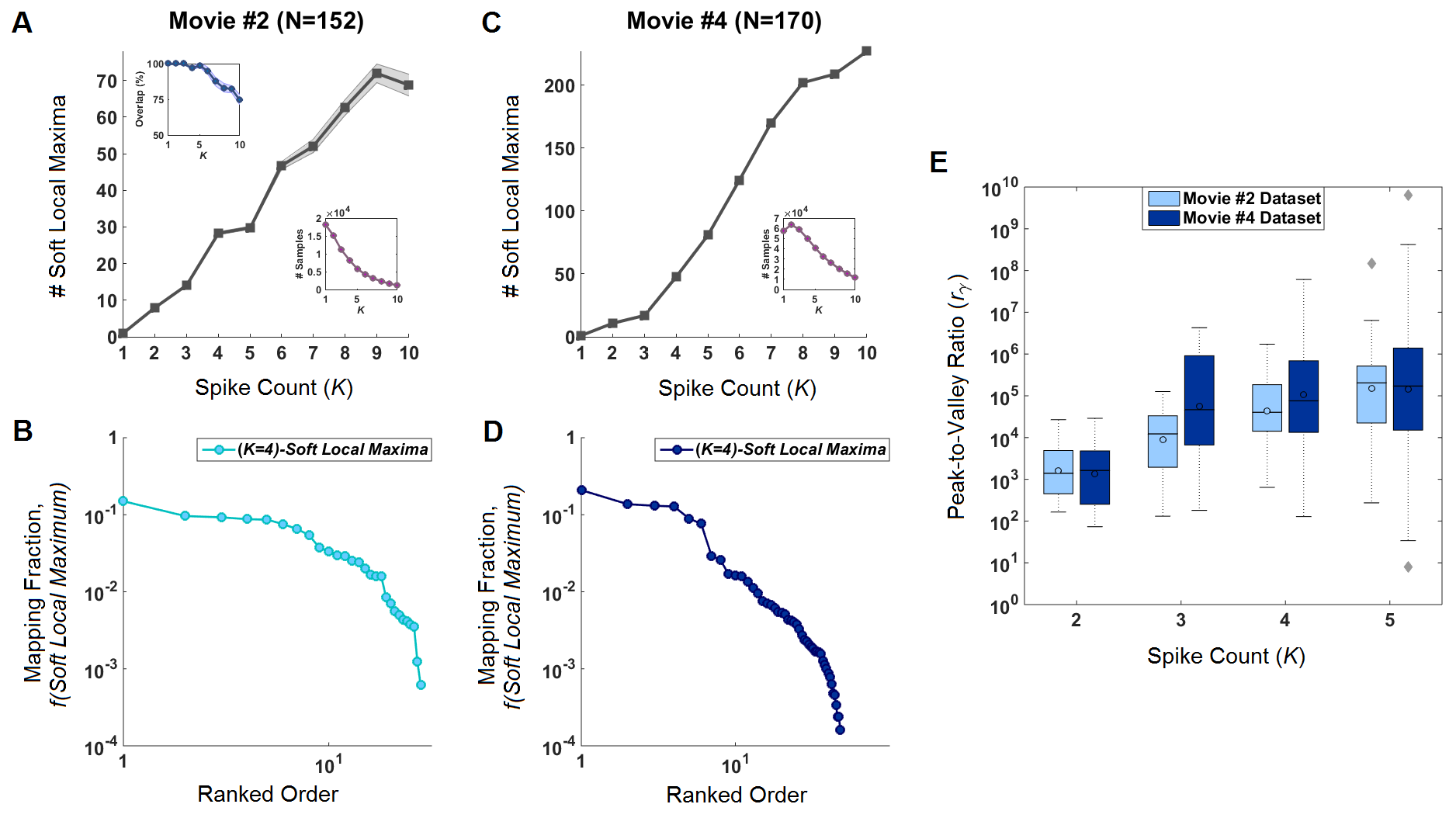}
\centering
\caption{Soft local maxima results for (A,B) the dataset of $152$ ganglion cells responding to Movie $\#2$; and (C,D) the dataset of $170$ ganglion cells responding to the original version of Movie $\#4$.   
(A) Shown in grey is the mean number of unique $K$-soft local maxima identified ($y$-axis) 
for each examined spike count level $K$, averaged over $100$ mapping iterations.  
Error bars denote one standard deviation over the $100$ iterations.  
\emph{Upper Left Inset}: The mean pairwise overlap ratio (see \emph{Appendix}) for each value of $K$ (blue), averaged over all 4950 iteration pairs.  Error bars denote one standard deviation.  
\emph{Lower Right Inset}: Total number of observed population responses in the data (purple) for each spike count level. 
(B) Example results for a specific spike count level $K$, arbitrarily chosen here to be $K=4$. Shown is a log-log plot of the proportion of the 8149 observed responses in the data with 4 spikes that were mapped  
to the corresponding $(K=4)$-soft local maximum indicated on the $x$-axis. 
(E) Box plot of the estimated peak-to-valley ratios of identified $K$-soft local maxima, denoted $r_{\vec{\gamma}}$, for each spike count level $K$ (see Appendix A).  
}
\label{fig6}
\end{figure*}

We next examined the responses of $170$ ganglion cells to the original version of Movie $\#4$ (described in section 2).  We similarly found a proliferation of soft local maxima for all examined spike counts, and observed that the number of identified unique soft local maxima monotonically increased as a function of spike count (Fig 6C).    

For both datasets, the distributions of the proportion of observed population responses that were mapped to each $K$-soft local maximum was nearly uniform for the top-ranked soft local maxima (Fig 6B,D).  
This is in stark contrast to the analogous local maxima results, in which $>99\%$ of observed population responses were mapped by single spin flip ascent to a single local maximum, the all-silent response (Fig 2A,C). 

We have shown that soft local maxima are present in the probability landscape of ganglion cell population responses under both non- and moderately-repeated visual stimulation.  But how pronounced is each soft local maximum in terms of its peak-to-valley ratio in the probability landscape?  
To quantify this, we computed a proxy measure that we denote $r_{\vec{\gamma}}$, which is a lower bound on the peak-to-valley ratio of a given soft local maximum $\vec{\gamma}$ (see Appendix A.7 for a formal definition).  
In general, we found that this ratio is quite high for nearly all soft local maxima, and moreover systematically increases with spike count (Fig 6E). 


\subsection{``Ridges" are a Feature of the Response Probability Landscape}  
So far we have restricted our attention to the set of all activity patterns having a fixed number of spikes, $K$.  
Within the disciplines of computer vision and differential geometry, there is a well-studied notion of ridges, which are curves or hypersurfaces composed of so-called ``ridge points" \citep{Haralick1983,EberlyRidges}.  
Although multiple definitions for ridge points exist, one popular definition is the \emph{height definition} \citep{EberlyRidges}.  
This definition stipulates that a necessary condition for a point $\vec{\sigma}$ to be a ridge point of a multivariate function $g$ is that it must be a \emph{generalized maximum}, which conceptually is a local maximum of $g$ when we restrict our search to a subspace of the function's domain.  
Although generalized maxima are technically only defined for continuous functions on vector space domains, we can see that the $K$-soft local maxima defined in Definition \ref{def:1} are a type of discrete analogue, where we have chosen the restricted metric subspace of the domain (which is the Hamming cube in our case) to be the set of all activity patterns with a fixed spike count, $K$.   

Motivated by this potential conceptual connection to ridge points, we next sought to investigate whether the soft local maxima found in the previous section are naturally organized \emph{across} spike count levels such that they comprise discrete ``ridges".  
Toward this purpose, we introduced a new definition, which we describe the intuition for in the following (see Appendix A.9 for the formal definition).  

\begin{figure}[!h]
\includegraphics[scale=0.35]{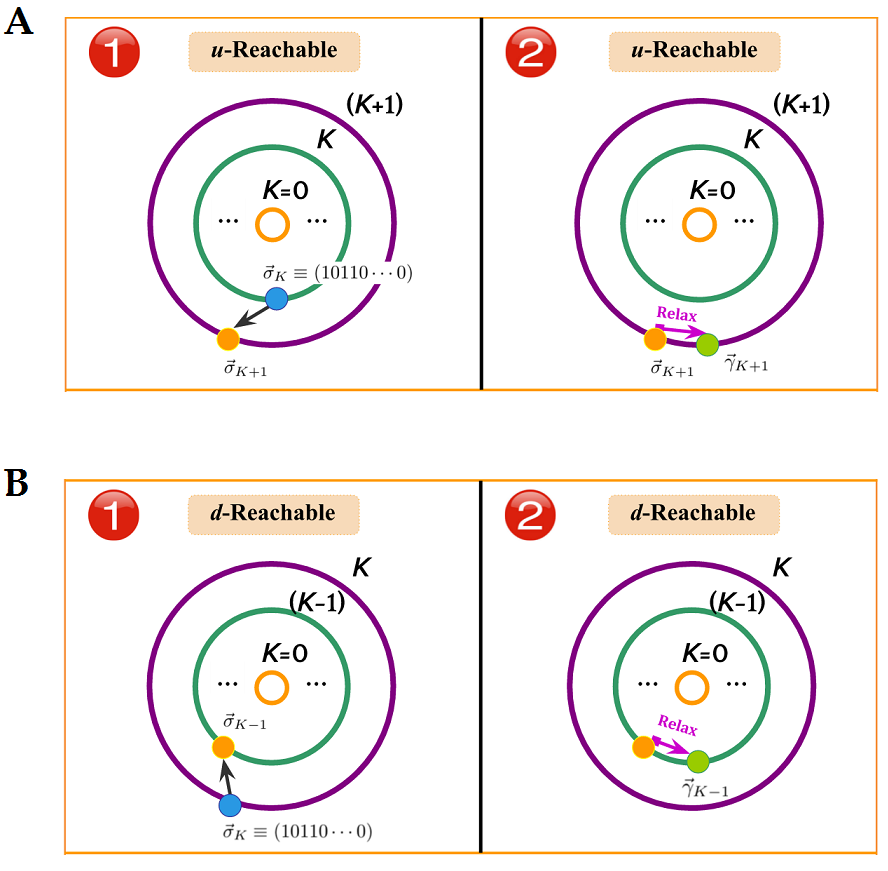}
\centering
\caption{Schematic illustrating the notions of \emph{u-reachable} and \emph{d-reachable}.   
(A) In this cartoon diagram, $\vec{\sigma}_K$ (blue dot) represents a $K$-soft local maximum, and $\vec{\gamma}_{K+1}$ (green dot) represents a ($K+1$)-soft local maximum. 
The different metric subspaces $\Omega_K$ of the full joint response space $\{0,1\}^N$, which are defined as $\Omega_K \equiv \{ \vec{\sigma} \in \{0,1\}^N \: | \: w_H(\vec{\sigma})=K \}$ and which we also refer to as ``spike count levels", are represented by colored concentric circles.  The center-most yellow circle represents the $0$th spike count level.  In this example, $\vec{\gamma}_{K+1}$ is $u$-reachable from $\vec{\sigma}_K$.  This is because:
(1) there exists a silent neuron $i$ such that changing neuron $i$'s instantaneous response to spiking (represented by the black arrow) will 
(2) result in a joint response pattern $\vec{\sigma}_{K+1}$ (orange dot) at the $(K+1)$-th spike count level that will be mapped via opposite sign neuron pair relaxation (represented by the purple arrow) to $\vec{\gamma}_{K+1}$. 
(B) In this cartoon diagram, the $(K-1)$-soft local maximum $\vec{\gamma}_{K-1}$ (green dot) is $d$-reachable from the $K$-soft local maximum $\vec{\sigma}_K$ (blue dot).  
}
\label{fig7}
\end{figure}

Intuitively, we want to explore whether there are natural `links' between soft local maxima across the different spike count levels.  If we focus on two spike count levels $K$ and $K+1$, then we can state this more concretely in terms of wanting to identify the $(K+1)$-soft local maxima that are `linked' in some intuitive way to a given $K$-soft local maximum of interest, $\vec{\sigma}_K$.   One natural way to think about being `linked' is in terms of being as `near' as possible, in terms of minimizing the number of computations needed to transition from $\vec{\sigma}_K$ to a $(K+1)$-soft local maximum.  
If the allowable computations are single spin flips (i.e. modifying one neuron's output activity) and opposite sign neuron pair relaxation, then the minimum number of computations possible to accomplish this transition is one implementation of each. 
This is because transitioning from $\vec{\sigma}_K$ to any activity pattern in the $(K+1)$-th spike count level will require at a minimum flipping one of $\vec{\sigma}_K$'s silent neurons to an active state.  
We let $\vec{\sigma}_{K+1}$ denote the activity pattern that differs from $\vec{\sigma}_K$ only in having one extra active neuron.  
Since $\vec{\sigma}_{K+1}$ is not guaranteed to be a soft local maximum, we must additionally allow for one implementation of opposite sign neuron pair relaxation applied to $\vec{\sigma}_{K+1}$, which will ensure transitioning to a $(K+1)$-soft local maximum, $\vec{\gamma}_{K+1}$.  
When it is possible to transition from a $K$-soft local maximum $\vec{\sigma}_K$ to a $(K+1)$-soft local maximum $\vec{\gamma}_{K+1}$ via the composition of one spin flip and one implementation of opposite sign neuron pair relaxation, then we say that $\vec{\gamma}_{K+1}$ is \emph{$u$-reachable} from $\vec{\sigma}_K$ (see Definition \ref{def:2} and Fig \ref{fig7}A).  
This formalism allows us to define a progression of `linked' soft local maxima spanning across low to high spike count levels.  
An analogous notion for exploring the organization of soft local maxima across high to low spike count levels is similarly defined (see Appendix A.9 and Fig \ref{fig7}B). 

To visualize how the soft local maxima are organized across low to high spike count levels, we used a standard class of search algorithm \citep{Skiena2008}.   We will here refer to our specific variant of this algorithm as the \emph{ridge search algorithm}, and we provide an intuition for how it works in the following (see Appendix A.10 and C for details). 
Our ridge search algorithm takes as input a single soft local maximum of interest, $\vec{\gamma}_{\text{root}}$.  
We will denote the spike count of $\vec{\gamma}_{\text{root}}$ by $K_{\text{min}}$, as this specifies the starting (and thus lowest) spike count level.  
In practice we chose $K_{\text{min}} = 4$, and chose each input to be one of the identified $(K=4)$-soft local maxima (section 3.3). 

Starting at $\vec{\gamma}_{\text{root}}$, the algorithm then proceeds to find all soft local maxima at the next higher spike count level, $K_{\text{min}}+1$, that are `linked' in terms of being $u$-reachable from $\vec{\gamma}_{\text{root}}$.  
We call the set of $(K_{\text{min}}+1)$-soft local maxima that are $u$-reachable from $\vec{\gamma}_{\text{root}}$ the \emph{neighborhood} of $\vec{\gamma}_{\text{root}}$.  
For each $(K_{\text{min}}+1)$-soft local maximum in the neighborhood of $\vec{\gamma}_{\text{root}}$, the algorithm then proceeds to find \emph{its} neighborhood of linked soft local maxima that reside in the next higher spike count level, $K_{\text{min}}+2$.  
This procedure is iterated recursively up to a specified maximum spike count level, $K_{\text{max}}$.  

In this way, our ridge search algorithm traces out all of the connection paths that start at the input soft local maximum $\vec{\gamma}_{\text{root}}$ and extend out to the highest specified spike count level. 
This information can be readily represented by a standard structure in graph theory called a \emph{rooted digraph} \citep{Thulasiraman1998}.  
We chose this graph theoretic representation to visualize the output of our ridge search algorithm (see Appendix A.10), because it is a $2D$ representation and hence inherently conducive to visualization, whereas the soft local maxima reside in a high-dimensional space.    
Shown in Figs 8 and 9 are example digraph visualizations that were obtained when we applied our ridge search algorithm to the dataset of $152$ ganglion cells responding to the non-repeated natural movie (Movie $\#2$).   

\begin{figure*}[!ht]
\includegraphics[scale=0.37]{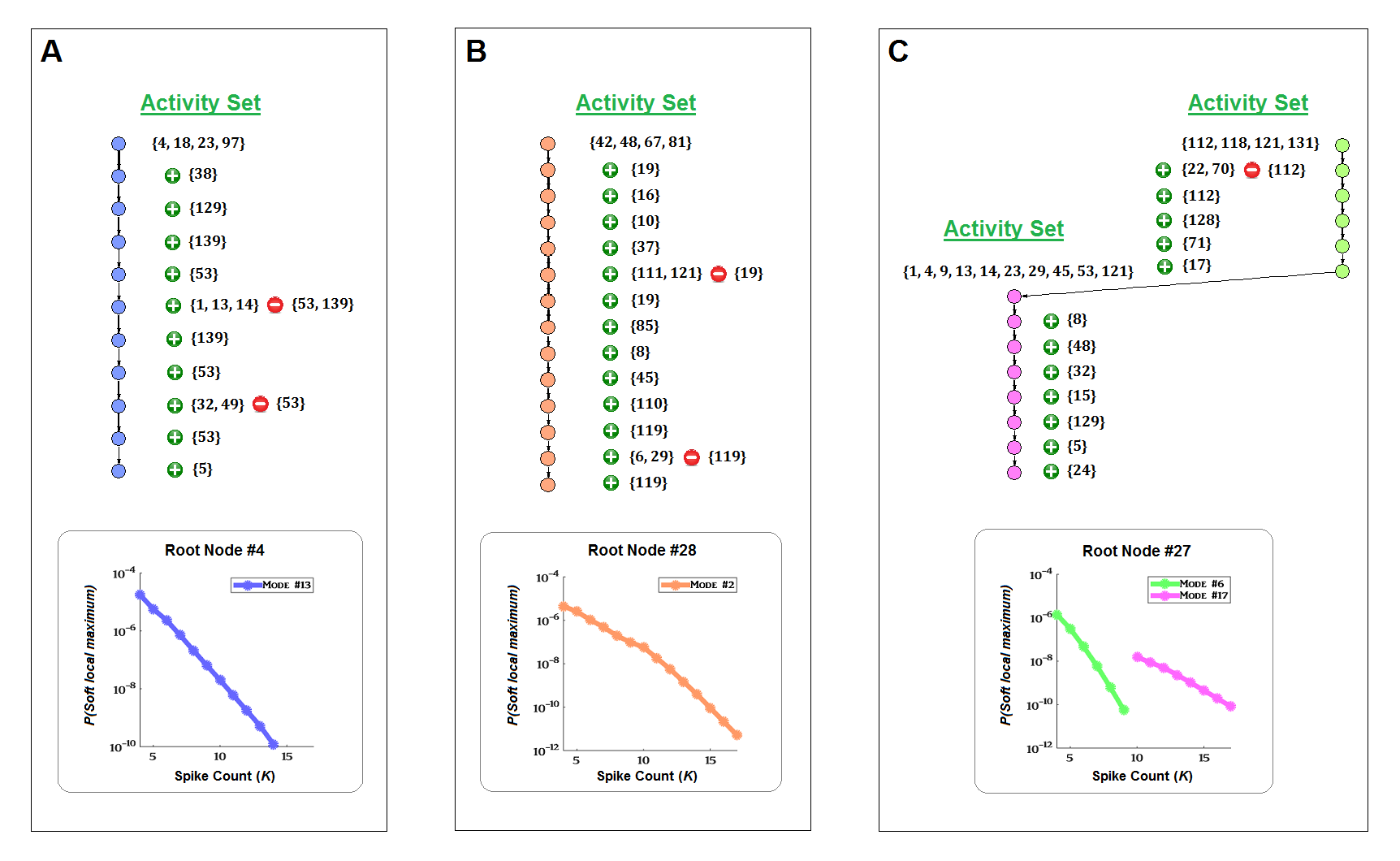}
\caption{Example Type 1 rooted digraphs obtained for the dataset of $152$ ganglion cells responding to Movie $\#2$.  
(A) Output digraph obtained via the ridge search algorithm when the unique ($K=4$)-soft local maximum with rank $4$ is the input.  
Labeled next to each node is the ``activity set" of the corresponding soft local maximum $\vec{\gamma}$, defined as $\mathcal{A}(\vec{\gamma}) \equiv \{\text{neurons} \: i \: | \gamma_i = 1 \}$.  
For ease of visualization, the activity set of the starting node is written out in full.  The activity set of each subsequent $(K+1)$-soft local maximum, $\vec{\gamma}_{K+1}$, is then denoted by the neuron(s) that is added to (green plus symbol) or removed from (red minus symbol) the previous $K$-soft local maximum's activity set to achieve $\mathcal{A}(\vec{\gamma}_{K+1})$.  
Node color denotes the MAP estimate of the associated collective mode.  
There was one distinct ridge for this example; all soft local maxima were associated with the same collective mode (blue). Below: Probability, as modeled by the Tree HMM, of each depicted soft local maximum vs. its spike count, $K$. 
(B,C) Same format as in panel (A), but for two other input ($K=4$)-soft local maxima. 
}
\label{fig8}
\end{figure*}

\begin{figure*}[!ht]
\includegraphics[scale=0.32]{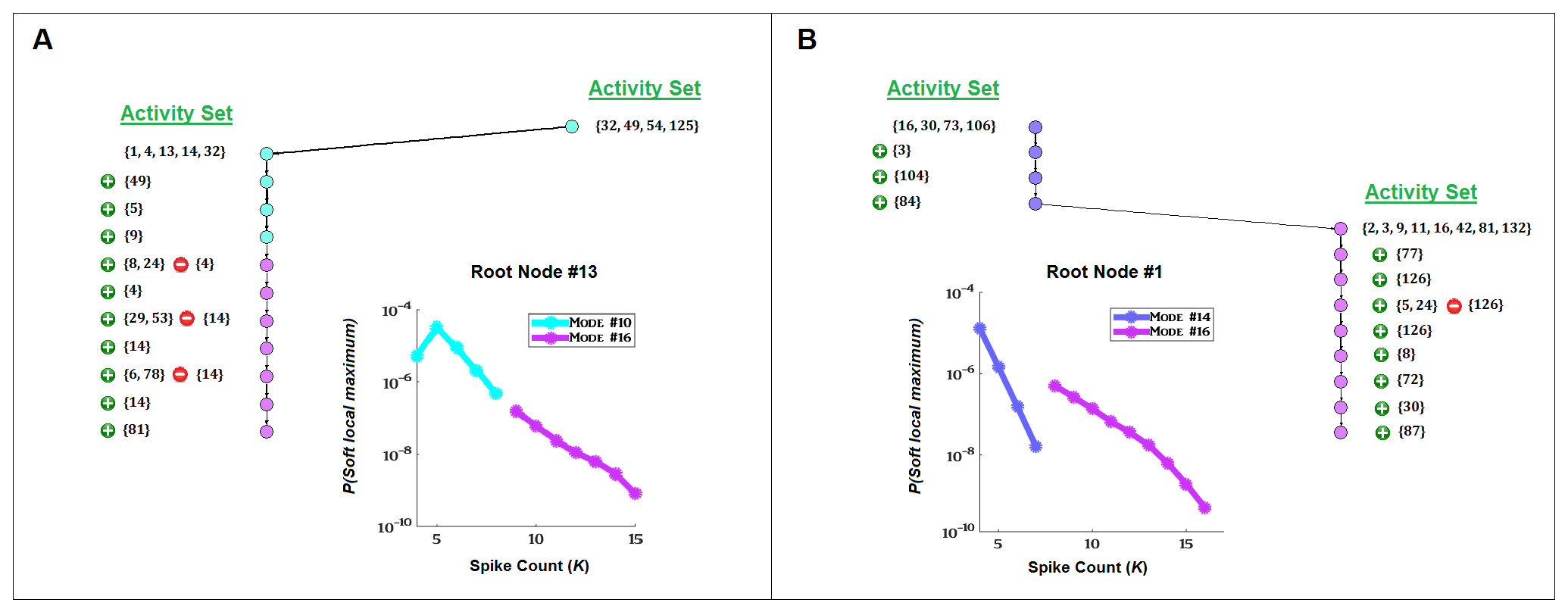}
\caption{Example Type 2 rooted digraphs obtained for the dataset of $152$ ganglion cells responding to Movie $\#2$.  
(A) Digraph obtained when the $(K=4)$-soft local maximum with rank $13$ was taken as the input to the ridge search algorithm.  
Notation is the same as in Fig 8. 
There was one distinct ridge; however, the soft local maxima comprising this ridge were associated with two possible collective modes: $\#10$ (cyan) and $\#16$ (violet). 
(B) Digraph obtained when the $(K=4)$-soft local maximum with rank $1$ was taken as the input.  
There were two distinct ridges. The first, which was comprised of soft local maxima that were all associated with the same collective mode ($\#14$, blue), was Type 1.  Soft local maxima comprising the second ridge were associated with collective mode $\#16$ (violet).  
Since mode $\#16$ corresponded with three distinct ridges, two of which are shown here in panels (A) and (B), this second ridge was classified as Type 2.
}
\label{fig9}
\end{figure*}

Since we can view soft local maxima as analogous to ridge points, this approach leads to an algorithmic definition of a discrete ridge.  
Specifically, a discrete ridge is the set of soft local maxima that are organized across low to high spike count levels such that each respective pair of $K$- and $(K+1)$-soft local maxima is `connected', in terms of being $u$-reachable.  
Note that in principle, the ridge search algorithm could trace out a single ridge starting at an input $\vec{\gamma}_{\text{root}}$ (as in the case when there is only one $u$-reachable soft local maximum at each successively higher spike count level), or could trace out multiple ridges. 
The number of ridges traced/visualized is a specific property of the dataset.  
We observed that many of the digraphs for the Movie $\#2$ dataset exhibited a single ridge (Fig 8A,B).  

In principle, it is also possible that a discrete ridge descending from $\vec{\gamma}_{\text{root}}$ could terminate at a lower spike count level $K_*$ than the arbitrarily-chosen $K_{\text{max}}$.  
An example occurrence of this scenario is shown in Fig 8C.  
To aid in visualizing distinct ridges, we further assigned an $x$- and $y$-coordinate to each node in the digraph visualizations.  
The intuition behind this $x$-coordinate assignment was to quantify whether or not there is a `jump' between ridges based on the amount of overlap between the different soft local maxima (see Appendix A.10 for formal details).  

\subsection{Collective Modes Correspond to Ridges}
For each soft local maximum $\vec{\gamma}$ in each digraph, we also computed the maximum \emph{a posteriori} (MAP) estimate of the collective mode that was most likely to be associated with it (see Eq. \ref{eq:14} in Appendix A.10).  To visualize this information, we then assigned a color to each node in the output digraphs, which uniquely identifies the associated collective mode.  
As seen in Fig 8, there was overall a strong tendency for each distinct ridge to be comprised of soft local maxima that were associated with the same collective mode. 

To broadly characterize the digraph visualization results in terms of the degree of correspondence between ridges and collective modes, we next classified each unique ridge as belonging to one of two disjoint categories: \emph{Type 1} or \emph{Type 2}.  
Formally, we classified a ridge as Type 1 if it corresponded with exactly one collective mode, which moreover uniquely corresponded with the ridge (that is, if there was a bijection between the ridge and a single collective mode).  
Conceptually, a Type 1 classification denotes that the given ridge has a `perfect' correspondence with one of the collective modes.   
We classified any ridge that was not Type 1 as Type 2.  
Note that we chose to classify on the basis of ridges, rather than on the basis of the digraph examples, to avoid potentially double-counting ridges when computing global statistics. 

Applying this classification scheme to the dataset of 152 ganglion cells responding to Movie $\#2$, we observed that 13 of the 17 identified unique ridges, i.e. $76.5\%$, were Type 1.   
Comparably, for the dataset of 170 ganglion cells responding to the original version of Movie $\#4$, we found that 32 of the 44 identified unique ridges, i.e. $73\%$, were Type 1.  
Hence for both a non-repeated as well as a moderately-repeated stimulus ensemble, we found that there was a substantial correspondence between the statistically-defined collective modes and the geometrically-defined ridges.  
In the remaining cases where there was not a `perfect' correspondence, either the ridge was associated with two collective modes (Fig. 9A), or a collective mode was associated with more than one ridge (Fig. 9B).  

\subsection{Ridges Correspond to ``Neuronal Communities"}
As shown in Fig 9, even in the few cases where there was not a perfect one-to-one correspondence between a given geometric ridge and a collective mode, intriguingly, we consistently found that each identified ridge corresponded with a specific group of active neurons.  
Specifically, for each ridge, which spans across multiple spike count levels, $K$, we observed that the active neuron sets of the soft local maxima comprising that ridge were nested (Figs 8 and 9).   
(By the ``active neuron set'' of a population response pattern $\vec{\gamma}$, we mean the set of neurons that have an instantaneous spiking response for that pattern, i.e. $\mathcal{A}(\vec{\gamma}) \equiv \{ \text{neurons} \:\: i \: | \: \gamma_i = 1 \}$).  
Stated another way, we found that population response patterns within each ridge exhibit active neurons that are members of an identifiable group of ganglion cells, combined with silence of all neurons outside of this group.  
We call the identifiable group of active neurons that is specific to each ridge the \emph{neuronal community} associated with that ridge.   

We next investigated how the full population of ganglion cells is organized in terms of these ridge-associated neuronal communities.  
To do this, for each ganglion cell $i$ (where $1 \leq i \leq N$) we recorded all of the neuronal communities it was a member of, based on the compiled digraph results from the previous section.   
We then visualized this information in the form of an undirected graph, a standard structure in graph theory \citep{Thulasiraman1998}.     
Since this graph visualizes the compilation of the ridge-associated neuronal community results for an entire dataset, we refer to it as the \emph{ridge union graph}. 
Each node in the ridge union graph represents a neuron in the population of ganglion cells, and an edge $(i,j)$ is present if ganglion cells $i$ and $j$ are members of at least one common neuronal community (see Appendix A.11 for details).   

The ridge union graph for the dataset of $152$ ganglion cells responding to the non-repeated natural movie is shown in Fig 10.   
As seen in Fig 10, mixed membership - that is, the case of a neuron having membership in more than one neuronal community - was prevalent.  
Specifically, averaged over the $83$ retinal ganglion cells that were assigned to at least one neuronal community, a given ganglion cell belonged to a mean ($\pm$ SEM) of $2.5 \pm 0.18$ neuronal communities. 
This high degree of mixed membership may be a signature of a combinatorial neural population code.

\begin{figure*}[!ht]
\includegraphics[scale=0.6]{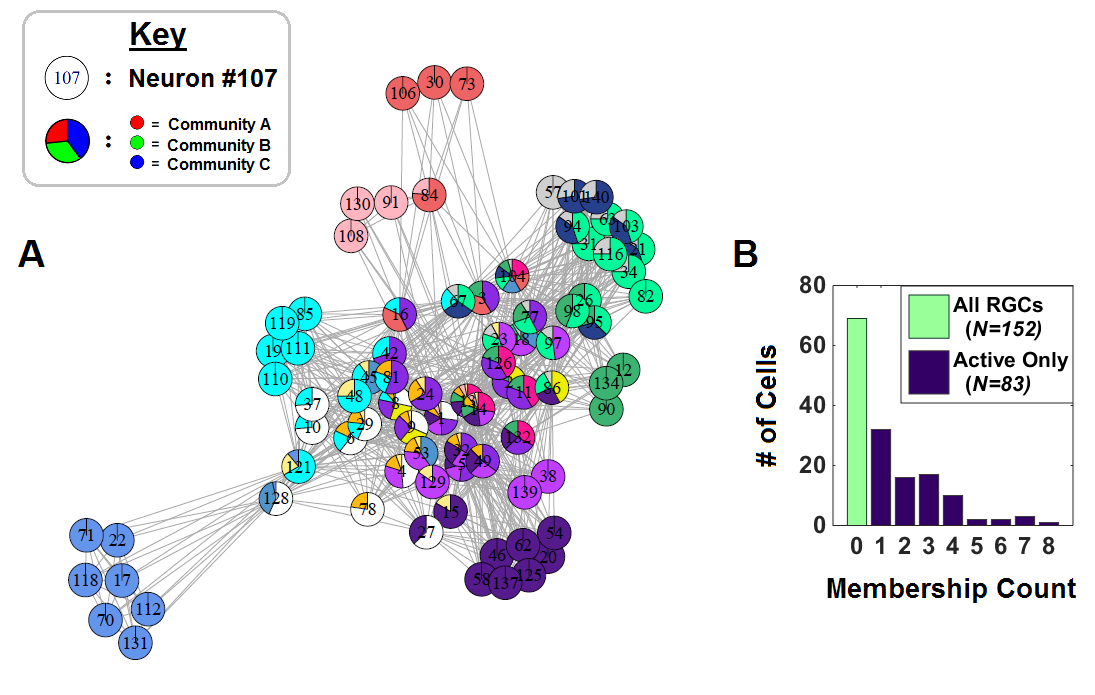}
\caption{Neuronal Communities. 
(A) Shown is the \emph{ridge union graph} for the dataset of $152$ ganglion cells responding to Movie $\#2$.  
Nodes represent individual ganglion cells in the population; each inscribed number indicates the index of the corresponding ganglion cell.  
The presence of an edge $(i,j)$ denotes that neurons $i$ and $j$ share at least one common ridge-associated neuronal community. 
Each color uniquely identifies one of the $17$ distinct ridge-associated neuronal communities that were identified for this dataset.  
Ganglion cells that exhibited mixed membership are depicted with ``pie" nodes (see Key).
(B) Histogram of the distribution of the degree of mixed membership across the population.   
Represented on the $x$-axis is the ``membership count", defined as the number of ridge-associated communities a given ganglion cell was a member of.  
Green denotes the same as purple, except that it also includes the $69$ ganglion cells that were not a member of any community.
}
\label{fig10}
\end{figure*}

In network science, \emph{community structure} refers to the occurrence of groups of nodes in a network that are more densely connected internally than with the rest of the network \citep{Fortunato2010}.   
Thus, another way to state our result is that we observed that the geometric picture of ridges in the joint response probability landscape naturally maps to one of communities in the population of ganglion cells.  
We performed two control analyses (see Appendix B for details) to verify that this organization of the ganglion cell population into many, relatively small neuronal communities is a specific property of the data, and not solely attributable to the underlying probability model.   

\section{Discussion}

\subsection{Error-Correcting Codes in Neuroscience} 
It is well established that noise is prevalent throughout the nervous system \citep{Mainen1995,Schneidman1998,Osborne2005,Faisal2008}.   
Partitioning noisy output patterns into fault-tolerant clusters corresponding with ``codewords" is a long-standing idea that has been utilized in traditional error-correcting codes in communications engineering \citep{Shannon1948}.   
In the engineering paradigm, a ``codeword" refers to the encoded form of some input information. The additional information (redundancy) added by the code prior to transmission to generate the codeword is used by a ``receiver" to enable the correction of errors induced by noise during transmission.  

Conceptually, we can draw an analogy between this paradigm and the retinal code as follows: the input information corresponds to the external visual stimulus (or everything within the bandwidth of the photoreceptors); the code that is used for encoding the external visual information into a form that can be utilized internally by the brain corresponds to the ganglion cell population code; and the population activity patterns that we as experimenters observe at the output of the ganglion cell layer correspond to potentially noise-corrupted versions of the codewords.   
However, we emphasize that there are also two fundamental disanalogies between the traditional engineering paradigm and our scenario: (1) in our case the original population code is unknown, as are its codewords; (2) we are here thinking about noise within the retina, not noise within a ``transmission line" that is separate from the input system.  
It is thus not conceptually possible in our case to characterize the noise in the ``transmission line".  

Since downstream brain areas lack direct access to external stimuli, biologically plausible decoding must be unsupervised. Error correction - that is, mapping noisy activity patterns to the correct codewords - therefore must be inherently unsupervised. If the population code were structured to enable the observed noisy activity patterns to be partitioned into clusters in the $N$-dimensional response space, such that each cluster corresponded to a codeword, then a major advantage would be that decoding and error correction could be performed simultaneously via an unsupervised clustering algorithm. Recent studies from multiple groups have suggested that neuronal population codes may indeed be qualitatively structured in this fashion \citep{Thesaurus2015,Tkacik2014,Prentice2016,Huang2016}.    
Note that there is a one-to-one mapping between codewords and clusters. Thus, since we lack explicit knowledge of the former, but can potentially detect the latter via unsupervised methods, we will henceforth (somewhat at variance with the notation in communications engineering) refer to the corresponding clusters as the neural population codewords. 

\subsection{Dependence of the Population Response Probability Landscape on Stimulus Ensemble Statistics} 
Arising from the maximum entropy literature has been the proposal that neuronal population codewords may correspond with local peaks in the joint response probability landscape \citep{Schneidman2006,Tkacik2014} (Fig 1). However, these studies analyzed population activity driven by many repeats of a short visual stimulus segment, which corresponds to a narrow range of variability in stimulus space. Here, we found that under low-repeat, i.e. broadly variable, stimulus ensembles, the joint response probability landscape was instead mostly devoid of local peaks (Figs 2 and 3).  

Why might the repeat structure of the stimulus ensemble affect the probability landscape so strongly? Our suggestion is that each stimulus elicits an average population response, and that repetition of the same stimulus produces scatter around that average response due to neural noise (Fig 11Ai). If the scatter due to noise (represented by black dots in Fig 11Ai) is small compared to the separation of the average response (red dot in Fig 11Ai) from responses evoked by other stimuli, then sampling over repeats will produce a local peak in the probability landscape. Such local peaks can be detected by an iterative hill climbing algorithm (Fig 11Aii). We expect that in the limit of many repeats of few stimuli, this scenario will be achieved. 

\begin{figure*}[!ht]
\centering
\includegraphics[scale=0.23]{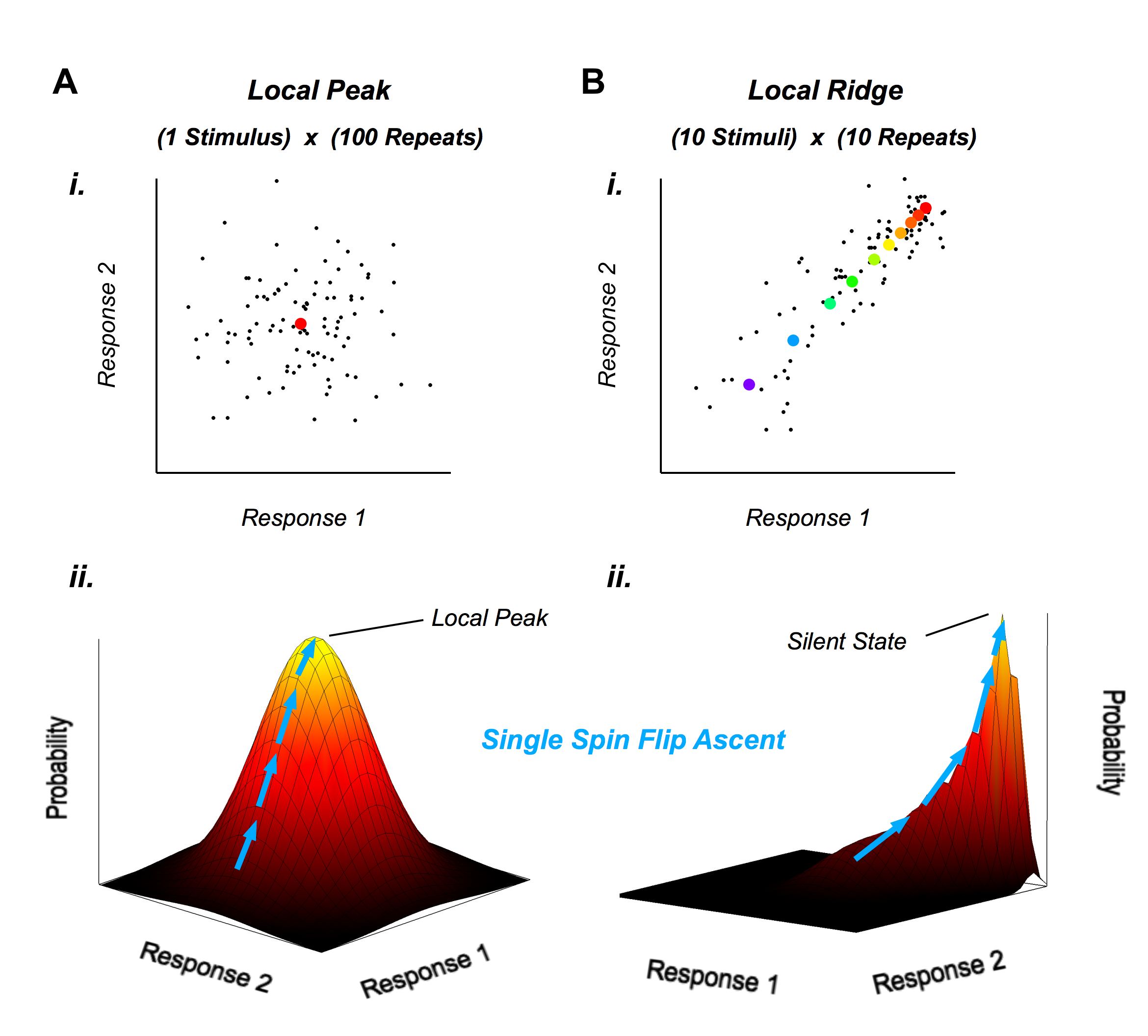}
\caption{ 
(A) \emph{i}. Example of neural responses shown in a 2D space triggered by 100 repeats of the same stimulus (black dots) along with the average response (red circle). \emph{ii}. The corresponding probability landscape contains a local peak, and single spin flip ascent maps neural activity to the local peak (blue arrows; each arrow represents one iteration comprising the ascent algorithm).  
(B) \emph{i}. Example of neural responses triggered by 10 repeats of 10 different stimuli (black dots) along with average responses (circles with different colors for each of the 10 stimuli). \emph{ii}. Here, the corresponding probability landscape is a local ridge, and single spin flip ascent maps neural activity to the all-silent state, regardless of spike count (blue arrows). 
}
\label{fig12}
\end{figure*}

However, in the case of a broader variety of stimuli, there is greater likelihood that average responses will be less well separated in response space. In this case, local peaks merge together (Fig 11B).  
In principle, this merging could simply result in fewer local peaks with negligible additional structure in the probability landscape. However, neural activity is often sparse, which causes there to be relatively more average population responses with low spike count than high. Due to this bias, the merged peaks can instead form a ridge. In this case, the same hill climbing algorithm will ascend to the all-silent state, which will be the global peak for sufficiently sparse neural activity (Fig. 11Bii). Notice that when the probability landscape is structured into ridges, the iterative single spin flip ascent algorithm can map neural activity states onto the all-silent global peak even if those states have high spike count. This is the result we found in our analysis (Figs 2 and 3).

\subsection{Comparing Two Principal Definitions of Population Codewords}
Energy basins (see Fig 1) and collective modes are candidate definitions of neural population codewords that arose independently from fundamentally different modeling frameworks.  
In particular, in past work that used the Maximum Entropy framework, energy basins in the modeled energy landscape were shown to exhibit error correction (that is, variable ganglion cell population responses to the same stimulus were robustly mapped by the single spin flip algorithm to the same local energy basin) for the highly-repeated stimulus regime \citep{Tkacik2014}.  
Likewise, using the hidden Markov model framework, it was recently shown that collective modes (which are the clusters of population responses that map to the same latent state of the hidden Markov model) exhibit error correction, and moreover provide a new feature basis set than ganglion cell receptive fields \citep{Prentice2016}.  
The explicit relationship between these candidate codewords was previously unknown, and is an empirical result of the present study (summarized in Table \ref{table1}).        
 
\begin{table*}[ht!]
\centering
  \caption{
 Comparison of Two Candidates for Population Codewords}
  \vskip 2mm
  \renewcommand{\arraystretch}{1.5}
	\begin{tabular}{c c c c c c}
	\cline{4-5}
	& & & \multicolumn{2}{ |L| }{\bf{Geometry in Probability Landscape}} & \\ \cline{1-6}
	\multicolumn{1}{ |M| }{\bf{Candidate Codeword}} & 
	\multicolumn{1}{ M| }{\bf{Model}} & 
	\multicolumn{1}{ M| }{\bf{Error-Correction Algorithm}} & 
	\multicolumn{1}{ M| }{\bf{High-Repeat Stimulus Ensemble}} & 
	\multicolumn{1}{ M| }{\bf{Low-Repeat Stimulus Ensemble}}  &  
	\multicolumn{1}{ M|  }{\bf{Correlate in Network of Neurons}} \\ \cline{1-6}
	\multicolumn{1}{ |M|  }{\bf{Energy Basin}} & 
	\multicolumn{1}{ M|  }{Maximum Entropy} & 
	\multicolumn{1}{ M|  }{Single spin flip ascent} & 
	\multicolumn{1}{ M|  }{Local peak} & 
	\multicolumn{1}{ M|  }{Absent} & 
	\multicolumn{1}{ M|  }{Unknown}     \\ \cline{1-6}
	\multicolumn{1}{ |M| }{\bf{Collective Mode}} &  
	 \multicolumn{1}{ M| }{Hidden Markov Model} & 
	\multicolumn{1}{ M| }{MAP estimation} & 
	\multicolumn{1}{ M|  }{Local peak} & 
	\multicolumn{1}{ M|  }{Ridge} & 
	\multicolumn{1}{ M|  }{Neuronal community}   \\ \cline{1-6}
	\end{tabular}
	\label{table1}
\end{table*} 

\noindent Understanding the relationship between these two concepts is important not just for the retinal code, but may be relevant for population codes throughout the brain. 
Our results here demonstrate that these concepts are not equivalent, and support the conclusion that collective modes are a better population codeword candidate, as they are robustly present in both the low-repeat and high-repeat stimulus regimes.  
Furthermore, we argue that both of these regimes are relevant in different behavioral contexts (see section 4.7).

\subsection{Other Structural Correlates of Collective Modes} 
It is important to emphasize that for both the Maximum Entropy and Tree hidden Markov models, the associated model fitting and error-correction algorithms (see Table 1) are non-biologically plausible computations.  
In order to explore biologically plausible algorithms that can learn the population codewords identified by the hidden Markov model (as we plan to do in subsequent work), it would be useful to understand whether the collective modes have other, potentially biologically-relevant structural correlates.  
The second main goal of this paper was to address this question.  
Our main results are that the statistically-defined collective modes (1) have the geometry of \emph{ridges} in the probability landscape of neural population activity (section 3.5); and (2) closely correspond to \emph{neuronal communities} within the population of ganglion cells (section 3.6), a notion from network science \citep{Fortunato2010}.  

\subsubsection{Properties of Neuronal Communities}
The mapping between the ridges (and hence collective modes) and neuronal communities arises from the following nontrivial property: for each given ridge, which spans across multiple spike count levels, we observed that the active neuron sets of the soft local maxima constituting that ridge were nested (Figs 8 and 9).  
Stated another way, we found that all soft local maxima within a given ridge exhibit active neurons that are members of an identifiable group of ganglion cells - which we call the \emph{neuronal community} associated with that ridge - combined with silence of all neurons outside of this group.  
Thus, the identity of the active and especially the silent neurons appears to be crucial in defining the population codeword.  
In contrast, the identity of the community is invariant to the precise number of active neurons above some threshold.  
Fundamentally, it is this invariance to the number of active neurons that gives rise to error correction. 

Note that formally, the community identities are sensitive to the highest spike count level we allow our ridge search algorithm to explore out to, $K_{\text{max}}$.  
However, due to the nested property and subsequent invariance of community identity to spike count, even if a given ridge extends to a higher spike count level $K_* > K_{\text{max}}$, the associated community we identify by exploring out to only $K_{\text{max}}$ will be very likely be conserved. 
Intuitively, if we think of community identity as the value of the ``angle" of neural activity in the response space, $\theta$ in Fig 5, then the nested property implies that a given ridge will have the same value of $\theta$ for $K_*$ and $K_{\text{max}}$.  
We therefore expect that the choice of $K_{\text{max}}$ will not qualitatively affect our results.  
Also, note that we have here enumerated only the ridges that start at spike count level $K_{\text{min}}=4$, or branch from a ridge starting at $K_{\text{min}}=4$.  
We would expect to find additional ridges (and thus communities) starting at higher $K_{\text{min}}$ values, as we hope to enumerate in future work.  The number of communities reported here is therefore a lower bound. 

\subsubsection{Biologically Plausible Decoding of Neuronal Communities}
Finally, if this type of structure plays a role in neural coding, then it would be important for downstream brain areas to be able to identify neuronal communities from population activity. In the present work, we determined community identity by starting with the ridge union graph for the ganglion cell population (Fig 10), and then using established simple community detection algorithms (see Appendix A.11).  
However, the downstream areas do not have access to this network information, but rather only the ganglion cell population activity.  

\begin{figure}[!h]
\centering
\includegraphics[scale=0.22]{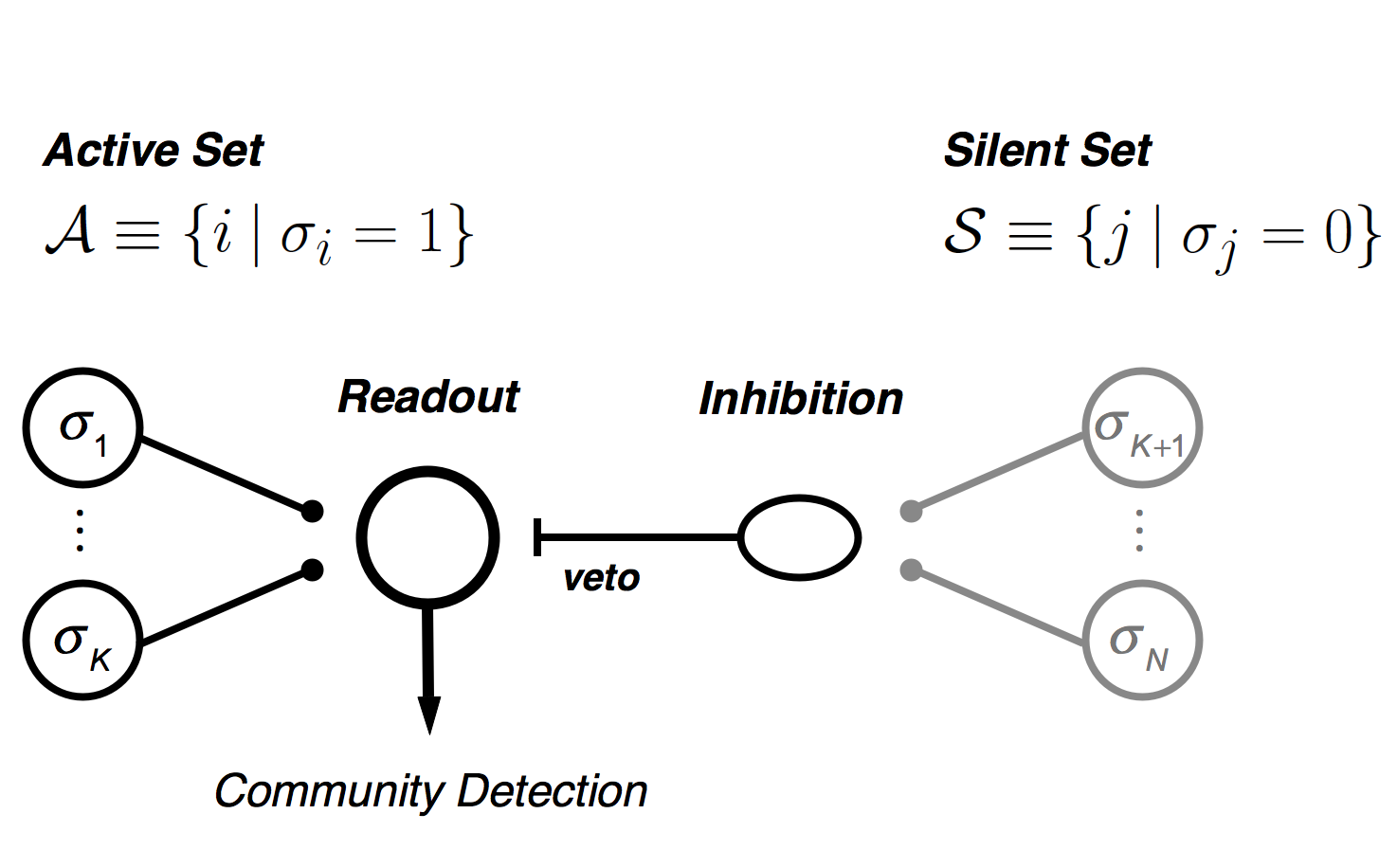}
\caption{Decoding Algorithm for Neuronal Communities. 
All of the $K$ neurons in the active set (black circles, left side) make excitatory synapses onto the readout unit (large circle).  The readout unit can be active as long as its input excitation is at least $K_{\text{min}}$.  All of the $(N-K)$ neurons in the silent set (grey circles, right) make excitatory synapses onto a local inhibitory interneuron (black oval).  
This neuron then feeds forward and vetoes activity in the readout unit.  This vetoing operation ensures that all members of the silent set must be silent in order for the readout unit to be active.
}
\label{fig13}
\end{figure}

Interestingly though, a simple, biologically plausible decoding algorithm exists for detecting neuronal communities.  
This algorithm (schematized in Fig \ref{fig13}) consists of feedforward excitatory and inhibitory synapses from the neural population onto a given readout unit, which will fire only if the community is present in its input population activity.  
For each member of the active set of neurons, the synapse should be excitatory.  
If, for example, each such synapse had the same weight, then the threshold of the readout unit would need to be $K_{\text{min}}$, and its output should saturate with just one spike for $K > K_{\text{min}}$.  
In this manner, the readout unit would be active if at least a criterion number of neurons in the active set of the community fire spikes.  

For each member of the silent set of neurons, there would then need to be a disynaptic pathway where activity of the neuron drives a local inhibitory cell, which feeds forward onto the readout neuron. If the inhibitory neuron had a threshold of one, then any member of the silent set could drive the inhibitory neuron to fire. Then, moreover, if that inhibitory neuron could strictly veto the firing of the readout unit - as is expected for chandelier-type inhibitory cells in the neocortex - this would enforce the condition that the readout unit would be silent if any of the silent set of input neurons fires. Finally, by making different choices of which cells have an excitatory synapse and which have a disynaptic inhibitory connection, different readout units would be selective for different neuronal communities within the same input neural population. 

\subsection{Connection to Donald Hebb's Cell Assemblies}
The organization of the ganglion cell population into a community structure with a high degree of mixed membership (Fig 10) is reminiscent of the concept of ``cell assemblies" hypothesized by Donald Hebb \citep{Hebb1949}.  
The first four defining properties of Hebb's original cell assembly concept are (paraphrased; see \cite{Sakurai1996,Hebb1949} for details): (i) overlapping set coding of information items; (ii) sparse coding; (iii) dynamic construction and reconstruction; and (iv) dynamic persistence. The neuronal community results found here are highly consistent with properties (i) and (ii). Specifically, we found that communities are overlapping sets of neurons: for the non-repeated natural movie dataset, a given ganglion cell was a member of on average $2.5 \pm 0.18$ different communities, and a given community overlapped with on average $10.7 \pm 0.59$ other communities. Moreover, any individual community contained a small subset (on average $12.1 \pm 1.2$ ganglion cells) of the 152 total neurons in the population, consistent with the sparse coding property. We cannot comment on how our neuronal community results relate to Hebb's properties (iii) and (iv), since these are dynamic properties, and the geometric ridges and consequently communities are features that were extracted from the modeled static probability landscape (Appendix A).

The neuronal community results obtained here are also consistent with the fifth and final property of Hebb's original cell assembly concept: the dynamic completion property, which stipulates that activation of a large enough subset of a cell assembly results in activation of the complete cell assembly \citep{Hebb1949}.  
Although intrinsic dynamics are not applicable in our case, there is a strong parallel between property (v) and our observation that population responses that had instantaneous activation of a large enough subset of the neuronal community were typically mapped to the same collective mode.  
It was shown in \cite{Prentice2016} that the collective modes exhibited a high degree of error correction, which combined with the present work suggests that communities likewise exhibit error correction.  
In combination, the three properties of overlapping coding sets, sparse coding, and robust completion were previously shown to offer theoretical advantages to overcome the deficiencies of the single-neuron doctrine \citep{Wickelgren1992,Legendy1967,Kanerva1988,Barlow1972}. 
Relatedly, in past work that modeled computation in visual cortex using groups of neurons called ``cliques" that satisfied the above three properties, it was derived that a circuit with this structure can take neurons that by themselves are crude and highly unreliable, and create aggregate units that are both extremely precise and highly reliable \citep{Zucker1999}.   

\subsection{Comparison to Other Definitions of Population ``Codewords"}
There have been many different proposals for how to study population neural codes. One particularly formative approach has been to apply information-theoretic techniques \citep{Rieke1999,Osborne2008,Schneidman2011,Brenner2000,Quiroga2009}.    
In past work that applied such information-theoretic methods to quantify combinatorial coding in the salamander and guinea pig retina, it was found that whereas synchronous spiking was mostly redundant, combinations of spiking and silence were generally synergistic \citep{Schneidman2011}.    
Our community results are consistent with and can be viewed as a generalization of this previous result to populations of $N > 100$ neurons. 

Another recent approach has been to investigate the possibility of clustering in the response space using a semantic distance metric \citep{Thesaurus2015}.    
In this work, it was shown that the studied 20-bit ganglion cell population responses nontrivially cluster in the response space based on semantic similarity.  
Unfortunately, we cannot directly compare the ridges or neuronal communities found here with the previously reported semantic clusters, since the approach used in \cite{Thesaurus2015} is a supervised approach that requires a high-repeat stimulus regime.  
In contrast, our analyses were applied to the non-repeated stimulus regime.  
Given our finding that the ganglion cell population response probability landscape is proliferated by local maxima for the high-repeat regime, we expect that semantic clusters will correspond geometrically to local maxima.  

However, we speculate that the probability landscape in the vicinity of such local peaks still has an extended, ridge-type geometry. The finding in \cite{Thesaurus2015} that simple linear and bilinear distance metrics do not correspond with semantic distance is thus consistent with our findings, as we would not expect simple linear or bilinear clustering routines to succeed well in capturing a ridge-type geometry.  
The reason for this intuition is that each ridge extends out from the all-silent state in a different ``direction" in the response space (see Fig 5), while linear and bilinear distance metrics use the same parameters everywhere on the probability landscape.  
Finally, our neuronal community results are also highly consistent with the finding that population responses belonging to the same semantic cluster exhibited a common subset of neurons that were always active, and other neurons that were always silent \citep{Thesaurus2015}.  


\subsection{Ridges vs. Peaks: Which are Present During Real Behavior?}
The fact that the geometry of the probability landscape of ganglion cell population activity is qualitatively different in the high-repeat versus low-repeat stimulus regimes raises an obvious question: which regime corresponds to the visual stimuli falling on the eye during real behavior? Answering this question depends on understanding on a more fundamental level how a ``stimulus ensemble" is defined from the viewpoint of the retina and downstream brain areas. In particular, there are two important considerations: (1) the duration of sampling of the stimulus space, and (2) whether readout processing is context-specific.

Regarding (1), during natural behavior readout circuits can potentially accumulate sampling over long time periods. In the case that sampling occurs over the entire lifetime of the organism, then the corresponding stimulus ensemble will be over all possible natural visual stimuli in the animal's environment. Because natural stimuli have been shown to possess conserved second-order statistics as well as other regularities like contrast scaling \citep{Dong1995}, this is a well-defined ensemble and such lifetime sampling can converge to a stable result.  
Consequently, for the retinal code in this limit, the stimulus ensemble would lack repeated features. Thus, for the case in which sampling occurs over the lifetime of the animal, we would expect that the probability landscape more closely resembles the ridge (i.e. non-repeated stimulus) regime studied here.

However, regarding (2), past work has shown that the neural code can be multiplexed, such that there is not necessarily one single interpretation extracted by downstream processing areas, but instead multiple context-specific interpretations that serve different purposes \citep{Victor1996,Victor1998,Meister1999,Fairhall2001}.     
It is possible that some downstream circuits that read out the ganglion cell population code could be gated by contextual signals, i.e. that these readout circuits only sample from the ganglion cell population under particular contexts.  
For example, an animal being in different environmental locations could constitute different contexts, signaled by feedback from the hippocampus; another example is running versus stationary movement, which is known to induce contextual modulation in V1 responses \citep{Ayaz2013}.  
These types of contextual signals could therefore restrict sampling to subsets of the visual environment in which some visual features repeat frequently.  
In this case, the probability landscape of ganglion cell population activity would have at least some local peaks that downstream circuits might benefit from identifying.  

There is another example of a context-dependent neural code that is highly relevant to neuroscience experiments.  
In many rigorously defined behavioral tasks, an animal is presented with two or several stimuli and required to use those stimuli to obtain a reward.    
In this case, the ``context" is a cue that initiates a behavioral trial, and/or the animal's presence in the behavioral rig.  
So if there are two salient visual stimuli in a given trial, then the retinal population code would likely have two well-separated peaks in this context.  
Similarly, population codes elsewhere in the cortex might also have two peaks in this context, especially after significant training.  
Furthermore, neural activity could very well be dense, rather than sparse, in this situation.  
For instance, if the context involves the sudden presentation of a salient stimulus, then many neurons could have a high probability of responding in this particular context.  
Because we believe that sparseness is necessary for the probability landscape to be organized into ridges, this factor also makes a peak-type probability landscape more likely.  
While one might argue that this kind of ``context" is artificial and not representative of natural animal behavior, this case will continue to be important in interpreting data in many neuroscience experiments.  

Within the scope of the full information processing system of the animal, these two regimes of sampling are not mutually exclusive.  
Accordingly, we speculate that the retinal population code contains at least two types of error-robust ``codewords".  
Namely, ridges may constitute one type that encode more general features such as classes of stimuli, whereas local maxima may constitute another type that encode more specific, individual stimulus features that occur in a given context.  

\subsection{The Generality of Ridge-Like Population Codes} 
There is no aspect of our approach that explicitly refers to properties of the retina: We made no assumptions about cells types, or receptive field properties of neurons, or functional models of feedforward sensory processing (such as the linear-nonlinear model) \citep{Rodieck1965,Chichilnisky2001}.  
In fact, nothing in our approach refers to visual processing or even sensory systems.  
The abstract nature of this approach therefore suggests that similar results might be found for population neural codes in many regions of the central brain.  


\newpage
\appendix 
\section{Appendix A: Analysis Details}

\subsection{A.1 Fitting the K-Pairwise Maximum Entropy Model}
The analytical form of the K-Pairwise MaxEnt distribution is \citep{Tkacik2014}:
\begin{eqnarray} \label{eq:1}
P^{(2,K)}\left(\vec{\sigma}\right) = \frac{1}{Z} \: e^{-\mathcal{H}(\vec{\sigma})}
\:,\: \text{where} \:\:\:
Z = \sum_{\vec{\sigma} \in \{0,1\}^N} e^{-\mathcal{H}(\vec{\sigma})}
\end{eqnarray}
\noindent and where the ``energy function" $\mathcal{H}$ is given by:
\begin{eqnarray} \label{eq:2}
\mathcal{H}\left(\vec{\sigma}\right) =
\sum_{i=1}^N h_{i} \sigma_{i} + \frac{1}{2} \sum_{i \neq j} J_{ij} \sigma_{i} \sigma_{j} + \sum_{K=0}^{N} \lambda_{K} \delta_{K, w_H(\vec{\sigma})}
\end{eqnarray}
\noindent where $\delta_{a,b}$ is the Kronecker delta function (i.e. $\delta_{a,b}=1$ if $a=b$ and $0$ otherwise), and $w_H(\vec{\sigma}) \equiv \sum_{i=1}^N \sigma_{i}$ is the Hamming weight of the population response $\vec{\sigma} \in \{0,1\}^N$.

\noindent We used the same learning procedure presented in \cite{Tkacik2014} to compute the parameters of the Hamiltonian for the K-Pairwise Maximum Entropy model given measured constraints. The proof of convergence for the core of this L1-regularized maximum entropy algorithm is given in \cite{Dudik2004}.  
The code used to fit the model was written in C++ and Matlab. 

\subsection{A.2 Additional Details for the Tree Hidden Markov Model}
Let $G_\alpha \equiv \Big( [N], \mathcal{E}_{\alpha} \Big)$ denote the underlying graph corresponding with emission distribution $Q_{\alpha}$ of the Tree HMM (i.e. the nodes of the graph $G_\alpha$ are the ganglion cell indices $[N] \equiv \{1,\cdots,i,\cdots,N\}$).  
Let $r_i$ denote the number of possible responses of neuron $i$.  
In practice, we allow the underlying graph to be a forest, so let $p$ denote the number of connected components.  Then the number of free parameters contributed by each $Q_{\alpha}$ (see Eq. \ref{eq:2}) is:
\begin{equation} \label{eq:4}
\begin{split} 
\text{\# of $Q_{\alpha}$ Parameters}  &= \sum_{(i,j) \in \mathcal{E}_{\alpha}} r_i r_j - \sum_{i \in [N]} \Big(\text{deg}(i) - 1\Big) r_i - p \\
&= \sum_{(i,j) \in \mathcal{E}_{\alpha}} r_i r_j - \sum_{i \in [N]}  r_i \cdot \text{deg}(i) \\ 
&\:\:\: + \sum_{i \in [N]} r_i - \underbrace{\left(N - \sum_{(i,j) \in \mathcal{E}_{\alpha}} 1 \right)}_p \\
&= \sum_{(i,j) \in \mathcal{E}_{\alpha}} (r_i - 1) (r_j -1) + \sum_{i \in [N]} r_i - N
\end{split} 
\end{equation}

\noindent In practice, we assume that all neuron responses are binary.  Thus, the total number of free parameters for the entire model is:
\begin{equation} \label{eq:5}
\begin{split}
\text{Total \# of Free Parameters} &= mN + m^2 + \sum_{\alpha = 1}^m | \mathcal{E}_{\alpha} | \\
&\leq m^2 + 2mN - m
\end{split}
\end{equation}

\noindent where $| \cdot |$ denotes cardinality. 
If the stationary distribution of the Markov chain is used, then the total number of free parameters reduces to $m(N+1) + \sum_{\alpha=1}^m | \mathcal{E}_{\alpha} | = \mathcal{O}(mN)$. 

For a fixed number of latent states $m$, the model was fit to data using the same learning procedure presented in \cite{Prentice2016}. In brief, the model parameters were inferred by maximum likelihood, using the Baum-Welch algorithm with an M-step modified to accommodate the tree graphical model form of the emission distributions $Q_{\alpha}$. Full details of the algorithm are described in \cite{Prentice2016}. The code used to fit the model was written in C++.

\subsubsection{A.2.1 Selecting the Optimal Latent Dimensionality}
To select the number of latent states $m$, also called the ``latent dimensionality" \citep{Lakshmanan2015}, we carried out a 2-fold cross-validation procedure in which we randomly chose half of the time bins in the data to assign to the training set.  
For the natural movie datasets, $m$ was chosen to be the value that maximized the cross-validated log likelihood (CV-LL), averaged over the 2 folds. 
Note that to mitigate overfitting, we also incorporated a regularization parameter $\eta \in [0,1]$ (in practice, $\eta = 0.002$ was used throughout), as in \cite{Prentice2016}. 
For the white noise checkerboard dataset in the high repeat regime, in practice the CV-LL curve often began to saturate at a lower latent dimensionality than the peak.  
In this case, as in \cite{Lakshmanan2015}, we report the latent dimensionality at which each CV-LL curve reached $90\%$ of its total height, where the height of each CV-LL curve is the difference between its maximum and minimum values over the range of dimensionalities tested. 
The motivation behind this choice is that it provides a more parsimonious representation of the data.

In \emph{Results} (section 3), we also report the \emph{normalized} cross-validated log-likelihood, $\widetilde{\mathcal{L}}$, which we define for a given latent dimensionality $m'$ as:
\begin{eqnarray} \label{eq:6}
\widetilde{\mathcal{L}} \left( m', \theta \right) = 
\frac{\big \langle \mathcal{L}(m',\theta) \big \rangle_n - \min_m \Big[ \big \langle \mathcal{L}(m,\theta) \big \rangle_n \Big]}{\max_m \Big[ \big \langle \mathcal{L}(m,\theta) \big \rangle \Big] - \min_m \Big[ \big \langle \mathcal{L}(m,\theta) \big \rangle \Big]}
\end{eqnarray}

\noindent where $\mathcal{L}(m,\theta)$ denotes the log-likelihood of the test dataset $D \in \{0,1\}^{N \times T/2}$ for the parameter values $\theta$ obtained via fitting the model to the training set, $\langle \cdot \rangle_n$ denotes the average over all $n$ cross-validation folds, and the minimum and maximum are taken over all latent dimensionality values $m$ considered.

\subsection{A.3 Parametric Repeat Analysis}
We here detail the design and implementation of the parametric repeat analysis reported in section 3.1.2, which aimed to investigate the relation between the repeat structure of the stimulus ensemble and prevalence of local probability maxima.  
As described in section 2, each of the two original visual stimulus ensembles used for this analysis was a movie (Movie $\#3$ or Movie $\#4$) that was designed to alternate between a presentation of a unique (i.e. non-repeated) movie segment, and a fixed ``target" movie segment. I.e. the unique movie segments and the repeated presentations of the target movie segment were interleaved. 
Derived from the original, full-length movie, we then generated a range of distinct shorter-duration movie stimuli that we refer to as ``subset movies".  Before detailing how these subset movies were constructed, we first introduce some notation: 
Let $A$ denote a unique movie segment, and let $B$ denote a repeated movie segment. 
We then define $n_A = $ the number of presentations of a unique movie segment, and define $n_B = $ the number of repeated presentations of the fixed target movie segment.  
We define $n_{\text{total}} = n_A + n_B$, and define the \emph{repeat ratio}, $\rho$, of the subset movie as $\rho = n_B/n_{\text{total}}$.  
Note that $0 \leq \rho \leq 1$. 

For the parametric repeat analysis for Movie $\#3$, we always set $n_{\text{total}} =70$ total movie segments comprising each subset movie, corresponding with a duration of $2100$ s (or 105,000 time bins).  
For each repeat ratio $\rho = n_B/n_{\text{total}}$ examined, we generated an associated subset movie by selecting $n_B$ of the repeated movie segments in the original Movie $\#3$. 
(For example, for a repeat ratio of $\rho = 0.1$, we chose $7$ repeat movie segments and $63$ unique movie segments to include in the associated subset movie).  
The specific choice of which $n_B$ of the $68$ repeat target segments to include, and which $n_A$ of the $70$ unique movie segments to include in the subset movie was made in accordance with one of five random sequence permutations. 
For the parametric repeat analysis for Movie $\#4$, we always set $n_{\text{total}} =72$ total movie segments comprising each subset movie, corresponding with a duration of $4320$ s (or 216,000 time bins).

Let $n_{\rho}$ denote the number of distinct repeat ratios examined, and let $n_k$ denote the number of random sequence permutations.  Then the total number of subset movies was $n_{\rho} \cdot n_k$.  
For each subset movie, we fit the Tree HMM to the population response data restricted to that subset movie. 
That is, from the original population spiking data, we selected the sequence of population response patterns that were elicited only by the sequence of movie segments included in the given subset movie. 
This restricted set of population spiking data corresponded to an $N \times 105000$ binary matrix for the analysis using Movie $\#3$ (or $N \times 216000$ matrix for the analysis using Movie $\#4$).  
The local maxima results reported in Fig 3 were obtained by then performing the single spin flip ascent algorithm on the joint response probability landscape corresponding with each subset movie (i.e. repeat ratio), as modeled by the Tree HMM.  

\subsection{A.4 Scaled Count Distribution}
To characterize the effect of spike count on the empirical joint response probability landscape (Fig 4), we defined the ``scaled count distribution", denoted $\widetilde{P}(K)$, as:
\begin{equation} \label{eq:8}
\widetilde{P}\left(K\right) \equiv \frac{ P_{\text{empirical}}(K) }{ \dbinom{N}{K} }
\end{equation}

\noindent where 
\begin{equation} \label{eq:9}
P_{\text{empirical}}(K) \equiv \frac{\sum_{t=1}^T \delta_{K, w_H\left(\vec{\sigma}(t)\right)}}{T}
\end{equation}

\noindent where $w_H \big( \vec{\sigma}(t) \big)$ denotes the spike count of the population response observed in time bin $t$, $T$ denotes the total number of time bins in the data, and $\delta_{a,b}$ is the Kronecker delta function (i.e. $\delta_{a,b}=1$ if $a=b$ and $0$ otherwise).

\subsection{A.5 Exploring the Modeled Probability Landscape}
\subsubsection{A.5.1 Finding Local Maxima} 
To find local probability maxima, we used the same algorithm as in \cite{Tkacik2014}, which we refer to as ``single spin flip ascent".  
Conceptually, this iterative algorithm is implemented as follows: 
Let $s$ index the current iteration step. 
The algorithm is initialized at $s=0$ by starting with a population response $\vec{\sigma} \in \{0,1\}^N$ in the data, which we denote as $\vec{\sigma}^{(0)} = \vec{\sigma}$. 
Within each iteration $s>0$, the algorithm then `flips' the response of neuron $i$, where $i$ is chosen in accordance with a random permutation of the $N$ neuron indices. 
(A different random permutation is independently generated for each iteration). 
That is, we set $\sigma_i^{(s)} = \sigma_i^{(s-1)} \oplus 1$, where $\oplus$ denotes mod 2 addition.  
The flip is retained if the new resulting configuration $\vec{\sigma}^{(s)}$ has lower energy (or equivalently, higher probability) than the previous iteration's pattern $\vec{\sigma}^{(s-1)}$, i.e. if $-\log P \big( \vec{\sigma}^{(s)} \big) < -\log P \big(\vec{\sigma}^{(s-1)}\big)$.  In this case, the algorithm continues to the next iteration step $s+1$. 
Otherwise, if flipping does not increase the probability, then the flip is not accepted, and the algorithm proceeds to try each of the other neuron indices according to the given permutation. 
When none of the $N$ neurons can be flipped, the resulting pattern is recorded as a local maximum.  Implementation of the single spin flip ascent algorithm was done in Matlab.

\subsubsection{A.5.2 Finding Soft Local Maxima} 
To find soft local maxima of the modeled probability landscape for a given spike count $K$ (see section 3.3), we used an iterative algorithm that we refer to as ``opposite sign neuron pair relaxation".  Conceptually, this algorithm is implemented as follows: 
Let $s$ denote the index of the current iteration step. 
The algorithm is initialized at $s=0$ by starting with a population response $\vec{\sigma}(t) \in \{0,1\}^N$ having $K$ spikes that appears in the data, denoted $\vec{\sigma}^{(0)} = \vec{\sigma}(t)$. 
Within each iteration $s>0$, the algorithm then randomly selects a pair of neurons $(i,j)$ that have opposite instantaneous responses according to the population response vector $\vec{\sigma}^{(s-1)}$, meaning that $\sigma_i^{(s-1)} = \sigma_j^{(s-1)} \oplus 1$, where $\oplus$ denotes mod $2$ addition.  
The algorithm then proceeds by `flipping' the response state of both neurons, by which we mean that we set $\sigma_i^{(s)} = \sigma_i^{(s-1)} \oplus 1$ and set $\sigma_j^{(s)} = \sigma_j^{(s-1)} \oplus 1$. 
In other words, the active neuron of the pair is made silent, and vice versa. 
This flipping alteration is retained if the resulting pattern, $\vec{\sigma}^{(s)}$, has a higher probability, i.e. if $P\left( \vec{\sigma}^{(s)} \right) > P \left( \vec{\sigma}^{(s-1)} \right)$. 
In this case, the algorithm continues to the next iteration step $s+1$. 
On the other hand, if flipping does not increase the probability, then the flipping alteration is not accepted, and the algorithm proceeds to try each of the other neuron pairs with opposite instantaneous responses according to a random permutation. 
If none of the $K \cdot (N-K)$ neuron pairs with opposite instantaneous responses can be flipped to increases the probability, then the algorithm terminates at iteration $s$. 
By Definition \ref{def:1}, the terminating pattern $\vec{\sigma}^{(s-1)}$ is a $K$-soft local maximum. 

\subsection{A.6 Assessing Robustness of the Set of Mapped Soft Local Maxima}
To investigate how robust the mapping from the dataset of all observed responses with $K$ spikes to the set of unique $K$-soft local maxima was, we performed $100$ independent implementations of our opposite sign neuron pair relaxation algorithm for the Movie $\#2$ dataset.  For each implementation, a distinct random sequence of permutations was used for the choice of neuron pairs when performing each iteration of the algorithm.  
For each value of $K$, we then computed the mean pairwise overlap ratio between the identified set of unique $K$-soft local maxima for each implementation pair $(l,m)$ (there was a total of $\binom{100}{2} = 4950$ implementation pairs), defined as:  
\begin{equation} \label{eq:10}
\text{Mean Pairwise Overlap Ratio} \equiv \frac{ \frac{1}{\binom{100}{2}} \sum_{l<m} \big| \mathcal{U}_l \cap \mathcal{U}_m \big| }{\max_{l=1}^{100} \big( \big| \mathcal{U}_l \big| \big)}
\end{equation}
 
\noindent where $l$ and $m$ index one of the 100 mapping implementations performed, $\mathcal{U}_l$ denotes the set of unique $K$-soft local maxima obtained for the $l$-th mapping implementation, and $| \cdot |$ denotes set cardinality.  Note that the the mean pairwise overlap ratio values are reported as percentages in Fig 6A. 

\subsection{A.7 Estimating Barrier-Depths of Soft Local Maxima}
To quantify how pronounced each identified soft local maximum $\vec{\gamma}$ was in terms of its associated peak-to-valley ratio in the probability landscape, we computed a proxy measure that we denote $r_{\vec{\gamma}}$.   Formally, this is defined as:
\begin{equation}
r_{\vec{\gamma}} \equiv \arg\max_{\vec{\sigma}_s} \left[ \frac{P\left(\vec{\gamma}\right)}{P\left(\vec{\sigma}_s\right)} \right] 
\end{equation}

\noindent where the argmax is taken over all populations responses $\vec{\sigma}_s$ observed in the data that are mapped (via opposite sign neuron pair relaxation) to $\vec{\gamma}$.  
Note that because our search is only over population responses $\vec{\sigma}_s$ observed in the real data, $r_{\vec{\gamma}}$ is a \emph{lower bound} on the true peak-to-valley ratio associated with soft local maximum $\vec{\gamma}$. 

To investigate the relationship between $r_{\vec{\gamma}}$ and spike count level, we performed the following analysis:  For each spike count level $K$ (in practice we examined $2 \leq K \leq 5$, because the set of identified $K$-soft local maxima was perfectly robust for this range), we computed the mean $r_{\vec{\gamma}}$, averaged over all $K$-soft local maxima.  
We also computed the interquartile range of the $r_{\vec{\gamma}}$ values for each spike count level $K$ (see Fig 6E).

\subsection{A.8 Choice of Probability Model}
We considered two models of the joint probability mass function (p.m.f.) of measured ganglion cell population activity: the K-Pairwise MaxEnt model and the Tree hidden Markov model \citep{Tkacik2014, Prentice2016}. For our purposes, it is important that the model accurately captures the soft local maxima present (or not) in the empirical probability landscape. 
Due to limited sampling, it is intractable to determine the full empirical probability landscape and hence to make a complete comparison.  
However, population responses with low spike count are far better sampled than those with high spike count.
We thus computed the empirical $(K=2)$-soft local maxima. 

This was done by performing $n$ iterations ($n=15$ in practice) of the following cross-validation procedure: For each iteration, we 
\begin{enumerate}
\item Randomly split the data (i.e. all observed population responses with 2 spikes) into two training sets.
\item For each of the two training sets, we estimated the empirical p.m.f. of an observed population response $\vec{\sigma}$ as:
\begin{equation} \label{eq:11}
\widehat{P}_{\text{empirical}}\left(\vec{\sigma} \right) = \frac{\sum_{t=1}^{T/2} \delta_{ \vec{\sigma}, \vec{\sigma}(t)} }{T/2}
\end{equation}
\noindent where $\vec{\sigma}(t)$ denotes the $t$-th observed population response, $T$/2 the total number of training samples, and $\delta_{a,b}$ the Kronecker delta function.

\item For each of the two training sets, we then performed opposite sign neuron pair relaxation on all unique observed responses in the given training half, using the respective estimated empirical p.m.f., to find $(K=2)$-soft local maxima. 
\end{enumerate}

\begin{figure*}[ht!]
\includegraphics[scale=0.4]{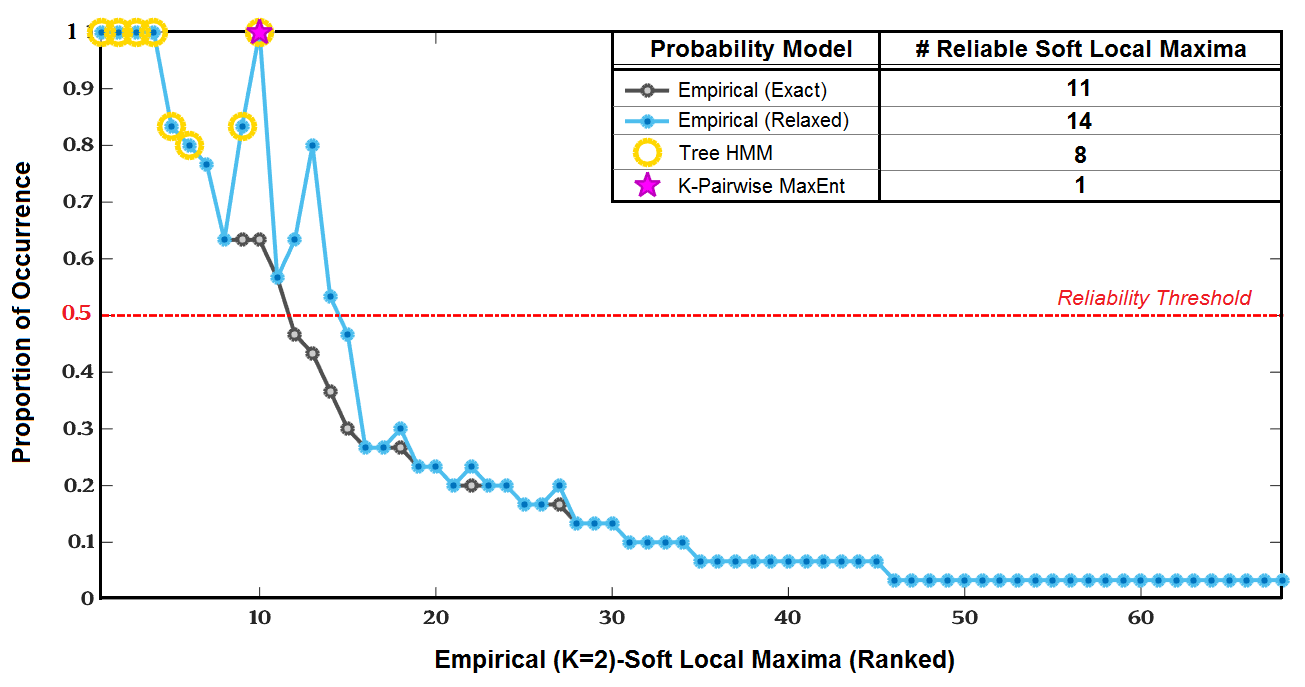}
\caption{Comparison of different models to the empirical (K=2)-soft local maxima results. 
Shown in grey/blue is the exact/relaxed proportion of occurrence of each of the 68 $(K=2)$-soft local maxima found using the halving procedure with the empirical probabilities (ranked).  
Soft local maxima found using the Tree HMM are denoted with a yellow annulus, and those found using the K-pairwise maximum entropy model are denoted with a pink star.  
The number of $(K=2)$-soft local maxima captured by each model with a proportion of occurrence greater than the chosen reliability threshold of $\Theta = 0.5$ is shown in the table inset.  
}
\label{fig14}
\end{figure*}

\noindent We then computed the union of all unique empirical $(K=2)$-soft local maxima found over the $2n$ iteration halves. Over $15$ iterations (i.e. $30$ different splits of the data), we found a total of $68$ unique $(K=2)$-soft local maxima. To check reliability, we then computed the proportion of occurrence of each unique $(K=2)$-soft local maxima, defined as:
\begin{equation} \label{eq:12}
\text{Proportion of Occurrence of} \:\: \vec{\gamma} := \frac{\sum_{h=1}^{n} \mathbbm{1}_{\mathcal{S}_h} (\vec{\gamma})}{2n}
\end{equation}
\noindent where $\vec{\gamma}$ denotes the soft local maximum of interest, $\mathcal{S}_h$ denotes the set of unique $(K=2)$-soft local maxima found after performing opposite sign neuron pair relaxation on the data in half $h$ (where $1 \leq h \leq 2n$), and $\mathbbm{1}$ denotes the indicator function (i.e. $\mathbbm{1}_{\mathcal{S}_h}(\vec{\gamma}) = 1$ if  $\vec{\gamma} \in \mathcal{S}_h$, 0 otherwise).  
Intuitively, the proportion of occurrence measures what fraction of all $2n$ data halves contain a given soft local maximum, $\vec{\gamma}$.
The resulting proportions of occurrence are shown in Fig \ref{fig14} (black trace).  

It is possible that differences obtained between the two splits of the data in the exact identities of the empirical soft local maxima could arise from noise, attributable to limited sampling upon halving the data.
To account for this possibility, we also performed a ``relaxed" version of the above empirical analysis. 

In the relaxed version, for each unique empirical $(K=2)$-soft local maxima found, we performed the following: For each of the $n$ iterations, we checked if the given $(K=2)$-soft local maximum - which we will call $\vec{\gamma}_1$ - was found when using one data half but not the other. If so, then we performed neuron pair relaxation on $\vec{\gamma}_1$, importantly using the p.m.f. estimated from the data in the second half. If this resulted in a $(K=2)$-soft local maximum that had been found originally in the second half - we will denote this $\vec{\gamma}_2$ - then we say that $\vec{\gamma}_1$ and $\vec{\gamma}_2$ are equivalent.  
Intuitively, we interpret $\vec{\gamma}_1$ and $\vec{\gamma}_2$ as being different only by sampling noise that shifted the local maximum in the first half, $\vec{\gamma}_1$, to a new activity pattern, $\vec{\gamma}_2$, in the second half of the data.
(In mathematical terms, we identify $\vec{\gamma}_1$ and $\vec{\gamma}_2$ as members of the same equivalence class).  
In this case, we updated the proportion of occurrence of $\vec{\gamma}_1$ by adding a term of $\frac{1}{2n}$. 
Otherwise, the proportion of occurrence of $\vec{\gamma}_1$ was unmodified. The updated proportions of occurrence for the relaxed empirical analysis are shown in Fig \ref{fig14} (light blue trace). 

In summary, the Tree HMM results matched the empirical results the best, capturing 8 of the reliable empirical $(K=2)$-soft local maxima (Fig \ref{fig14}).  
In contrast, the K-Pairwise MaxEnt model captured only one $(K=2)$-soft local maximum.

\subsection{A.9 Searching for Ridges} 

We sought to investigate the organization of soft local maxima \emph{across} low to high spike count levels.  Toward this purpose, we introduced the following definition: 
\begin{definition} \label{def:2}
A soft local maximum $\vec{\gamma}$ is \emph{u-reachable} from another soft local maximum $\vec{\sigma}$ if and only if $w_H(\vec{\gamma}) = w_H(\vec{\sigma}) + 1$, and $\exists$ neuron $i \in$ the set of neurons that have a silent response, $\mathcal{S}(\vec{\sigma})$, such that:
\begin{itemize}
	\item[$(i)$] Performing opposite sign neuron pair relaxation on $\vec{\sigma}^{(i)}$ results in $\vec{\gamma}$, and  
	\item[$(ii)$] $P[\vec{\sigma}^{(i)}] < P[\vec{\sigma}]$  (to ensure that $\vec{\sigma}^{(i)}$ is not a local maximum)
\end{itemize}
\noindent where $w_H(\cdot)$ denotes spike count, $\vec{\sigma}^{(i)}$ denotes the population response pattern that differs from $\vec{\sigma}$ only in switching neuron $i$'s state from silent to active, and $\mathcal{S}(\vec{\sigma}) \equiv \{ \text{neurons} \: j \: | \: \sigma_j = 0 \}$. 
\end{definition}

\noindent We also introduced the following analogous definition for investigating the organization of soft local maxima across high to low spike count levels:
\begin{definition} \label{def:3}
A soft local maximum $\vec{\gamma}$ is \emph{d-reachable} from another soft local maximum $\vec{\sigma}$ if and only if $w_H(\vec{\gamma}) = w_H(\vec{\sigma}) - 1$, and $\exists$ neuron $i \in$ the set of neurons that have an active response, $\mathcal{A}(\vec{\sigma})$, such that:
\begin{itemize}
	\item[$(i)$] Performing opposite sign neuron pair relaxation on $\vec{\sigma}^{(i)}$ results in $\vec{\gamma}$, and  
	\item[$(ii)$] $P[\vec{\sigma}^{(i)}] > P[\vec{\sigma}]$  
\end{itemize}
\noindent where $\vec{\sigma}^{(i)}$ denotes the population response pattern that differs from $\vec{\sigma}$ only in switching neuron $i$'s state from active to silent, and $\mathcal{A}(\vec{\sigma}) \equiv \{ \text{neurons} \: j \: | \: \sigma_j = 1 \}$. 
\end{definition}

\subsection{A.10 Visualizing Ridges} 
To visualize specific examples of how soft local maxima were organized across low to high spike count ($K$) levels, we used a type of breadth-first search (BFS) algorithm \citep{Skiena2008}.    
We call our variant of this algorithm the \emph{ridge search algorithm}.  
The input to the ridge search algorithm is a given ``root" soft local maximum, $\vec{\gamma}_{\text{root}}$, and the output is a rooted digraph that is specific to the input.  
The nodes of this output rooted digraph represent soft local maxima, and a directed edge $(\vec{\gamma}_K,\vec{\gamma}_{K+1})$ is present if and only if soft local maximum $\vec{\gamma}_{K+1}$ is $u$-reachable from $\vec{\gamma}_K$, as defined in Definition \ref{def:2}.  
Given a fixed spike count level $K_{\text{min}}$, we chose each input to be one of the $K_{\text{min}}$-soft local maxima found by performing opposite sign neuron pair relaxation on the probability landscape obtained by fitting the Tree HMM to the data.  

Note that the ridge search algorithm is non-deterministic, since one of its component subroutines is the non-deterministic opposite sign neuron pair relaxation algorithm.  
To incorporate our confidence level about the digraph edges computed by the ridge search algorithm, we thus did the following:  For each neuron $i$ in the silent set of the current $K$-soft local maximum, $\mathcal{S}(\vec{\gamma}) \equiv \{ \text{neurons} \: j \: | \gamma_j = 0 \}$, we performed $10$ iterations of opposite sign neuron pair relaxation on the population response pattern in the $(K+1)$-th spike count level that was obtained when neuron $i$'s response was changed to spiking.  For each iteration, we used a different, independent choice of the random permutation.  
We then set the weight of each directed edge $\big( \vec{\gamma}, \vec{\gamma}_{K+1} \big)$, which we denote by $w(\vec{\gamma}, \vec{\gamma}_{K+1})$, as the proportion of the $10$ relaxation iterations that resulted in $\vec{\gamma}_{K+1}$.    
To mitigate effects due to noise, for each unique $(K+1)$-soft local maximum $\vec{\gamma}_{(K+1)}$ reached after performing the above procedure on a given $\vec{\gamma}$, we further only included the directed edge $\big( \vec{\gamma}, \vec{\gamma}_{K+1} \big)$ if $w(\vec{\gamma}, \vec{\gamma}_{K+1})$ exceeded a reliability threshold, $\Theta$.  
In practice, we used $\Theta = 0.3$.  
The computed weight of each edge is represented by edge thickness in the digraph visualizations (Figs 8 and 9). 

To aid with visualizing distinct ridges, we also assigned an $x$- and $y$-coordinate to each node $\vec{\gamma}_K$ (i.e. soft local maximum) in the output rooted digraphs as follows:
\begin{enumerate}
\item The $y$-coordinate, denoted $y(\vec{\gamma}_K)$, was set to $K$ (i.e. the spike-count of $\vec{\gamma}_K$).
\item Our choice of formulation for the $x$-coordinate was based on two motivating criteria:  First, we wanted to be able to visualize potential ``jumps" in ridge organization.  These jumps can arise due to the current identified ridge either terminating at a lower spike count level than the one we chose arbitrarily to explore out to, or branching into multiple distinct ridges.  
Examples of this scenario can be seen in Figs 8C and 9B.   
Second, we wanted to visualize only ``genuine" jumps or branching, with respect to the notion of neuronal communities.  
Specifically, after inspection of preliminary digraph visualizations, we noticed that `connected' (i.e. $u$-reachable) soft local maxima typically had a set of active neurons that was a subset of a larger unique group (which we call the \emph{neuronal community}).  
We thus post-hoc formulated our definition of the $x$-coordinate for nodes in our digraph visualizations to aid us in examining how prevalent this property was.  

To do this, we needed to define an overlap measure between the activity sets of different soft local maxima.  The natural distance metric based on overlap is:
\begin{eqnarray} \label{eq:13}
d_O\left(\vec{\gamma}_K,\vec{\gamma}_G\right) \equiv | \mathcal{A}\left(\vec{\gamma}_K\right) \cap \mathcal{A}\left(\vec{\gamma}_G\right) | - K
\end{eqnarray}

\noindent where $G>K$ denotes a higher spike-count level, $\mathcal{A}(\vec{\gamma}) := \{ i | \gamma_i = 1\}$ denotes the ``activity set" (i.e. the set of active neurons) of soft local maximum $\vec{\gamma}$, and $| \cdot |$ denotes set cardinality.  
Note that if the activity set of a $K$-soft local maximum is a subset of the activity set of a soft local maximum at a higher spike count level, $G>K$, then this distance is $0$.  I.e. if $\mathcal{A}(\vec{\gamma}_K) \subset \mathcal{A}(\vec{\gamma}_G)$, then $| \mathcal{A}\left(\vec{\gamma}_K\right) \cap \mathcal{A}\left(\vec{\gamma}_G\right) | = K$, and thus $d_O\left(\vec{\gamma}_K,\vec{\gamma}_G\right)=0$. 

The input to our procedure is the digraph information - that is, the set of nodes (soft local maxima) and the set of edges connecting the nodes; the output is the assigned $x$-coordinate of each node.  
In brief, our procedure is initialized by setting the $x$-coordinate of the soft local maximum at the highest spike count level, which we denote $\vec{\gamma}_{K_{\text{max}}}$, to $0$.  That is, in our notation, we set $x(\vec{\gamma}_{K_{\text{max}}})=0$.  
Our procedure then works backward to compute the $x$-coordinate of each node at a successively lower spike count level.  
Specifically, for each node $\vec{\gamma}_K$ where $K_{\text{min}} < K < K_{\text{max}}$, we compute its \emph{path neighborhood}, $\mathcal{N}_p(\vec{\gamma}_K)$, which is the set of all soft local maxima $\vec{\gamma}_G$ residing in a higher spike count level $G>K$ such that there is a path in the digraph connecting $\vec{\gamma}_K$ and $\vec{\gamma}_G$.   
If the path neighborhood is empty, i.e. $\mathcal{N}_p(\vec{\gamma}_K) = \emptyset$, then this means that $\vec{\gamma}_K$ is situated at the end of a distinct branch.  
In this case, we assign $x(\vec{\gamma}_K) = x(\vec{\gamma}_{K_{\text{max}}}) + b$, where $b$ is an offset. 
Otherwise, if $\mathcal{N}_p(\vec{\gamma}_K) \neq \emptyset$, then we found the soft local maximum in the path neighborhood that was closest to $\vec{\gamma}_K$ in terms of overlap, which we denote $\vec{\gamma}_G^*$.  Finally, we assigned $\vec{\gamma}_K$ the same $x$-coordinate as this nearest $\vec{\gamma}_G^*$, plus an offset equal to the overlap distance between $\vec{\gamma}_K$ and $\vec{\gamma}_G^*$: 
\begin{eqnarray} \label{eq:14}
x\left( \vec{\gamma}_K \right) = x\left( \vec{\gamma}_G^* \right) + d_O \left(\vec{\gamma}_K, \vec{\gamma}_G^* \right) \\
\text{where} \:\:\:\: 
\vec{\gamma}_G^* \equiv \argmin_{\vec{\gamma}_G \in \mathcal{N}_p(\vec{\gamma}_K)} d_O \left(\vec{\gamma}_K, \vec{\gamma}_G \right)
\end{eqnarray}

\noindent In the case of a tie (e.g. if a $(K=4)$-soft local maximum $\vec{\gamma}_4$ had complete overlap of its activity set with both $\vec{\gamma}_5$ and $\vec{\gamma}_6$), we chose the soft local maximum at the nearest spike count level to be $\vec{\gamma}_G^*$. (I.e. in the above example, we would choose $\vec{\gamma}_G^* = \vec{\gamma}_5$, and would assign $x\left(\vec{\gamma}_4\right) = x\left(\vec{\gamma}_5\right)$). 


\end{enumerate}

\noindent In addition, we computed the MAP (maximum \emph{a posteriori}) estimate $\alpha$ of the Tree HMM latent state for the soft local maximum $\vec{\gamma}$ associated with each digraph node:
\begin{equation} \label{eq:16}
\begin{split}
\hat{\alpha}_{\text{MAP}} (\vec{\gamma}) &\equiv \arg\max_{\alpha}  P\left(\alpha | \vec{\gamma}\right) \\
&= \arg\max_{\alpha} \left[ \frac{P(\vec{\gamma}|\alpha) P(\alpha)}{\int_{\beta} P(\vec{\gamma}|\beta) P(\beta) d\beta} \right] \\
&= \arg\max_{\alpha} \left[ \psi_{\alpha} Q_{\alpha}(\vec{\gamma}) \right]
\end{split}
\end{equation}

\noindent where $\psi$ denotes the stationary distribution of the Markov chain, and $Q_{\alpha}(\cdot)$ is the emission distribution for mode $\alpha$ (see Appendix A.2).    
Visualization of the rooted digraphs computed via the ridge search algorithm (examples of which are shown in Figs 8 and 9) was automated using the \emph{igraph} and \emph{network} packages in R. 

We also investigated the organization of soft local maxima across high to low spike count ($K$) levels; example results are shown in Appendix D. For this analysis, a rooted digraph was constructed, but where each root node was taken to be a soft local maximum found at a high $K$ level, and where a directed edge $(\vec{\gamma}_K,\vec{\gamma}_{K-1})$ was added if and only if soft local maximum $\vec{\gamma}_{K-1}$ is $d$-reachable from $\vec{\gamma}_K$ (see Definition \ref{def:3}).  
Note that we report the $u$-reachable version (i.e. starting from a low spike count and progressing out to higher spike count levels) in the main text, as we have more reliable sampling of the data at low spike count levels.

\subsection{A.11 Constructing the Ridge Union Graph} 
In network science, \emph{community structure} refers to the occurrence of groups of nodes in a network that are more densely connected internally than with the rest of the network \citep{Fortunato2010}.   
In parallel with this concept, we observed that we can map the geometric picture of ridges in the joint response probability landscape to one of communities in the population of ganglion cells: We start with the empty graph (i.e. no edges) in which each node represents a ganglion cell in the population. For each ridge, we then compute the union of the set of active neurons for each soft local maximum comprising that ridge.  
This union set, which we call the \emph{neuronal community} associated with the ridge, is then represented as a clique in the undirected graph.  That is, we add all-to-all connectivity in the graph between the neurons in the union set. When we have added in the clique corresponding with each ridge, we call the resulting network the \emph{ridge union graph} (see Fig 10A).  

The inverse process of identifying the neuronal communities given the ridge union graph can be readily implemented by simple community detection algorithms. 
Specifically, all neuronal communities can be identified in this case via the Bron-Kerbosch algorithm, which is a well-known algorithm for finding the maximal cliques in an undirected graph \citep{Bron1973}.    
Construction of the ridge union graph and simple community detection (see Fig 10A) was done using the \emph{igraph} package in R.

\section{Appendix B: Control Analyses}
We sought to ascertain that correlations in the data were necessary to give rise to the soft local maxima and discrete ridge structures observed here, as opposed to these features trivially arising from the underlying probability model.    

\subsection{B.1 Heterogeneous Firing Rates Shuffled Control}
To generate this control, for each neuron $i$ in the population we performed a ``complete shuffle" on neuron $i$'s discretized spike train (binned into $T$ total $20$ ms time bins) occurring in the data, by implementing a random permutation of the $T$ time bins. 
Note that an independent random permutation was implemented for each neuron. This complete shuffling procedure eliminates all signal and noise correlations among neurons in the population, but retains the firing rate of each neuron over the course of the experimental recording. 
Thus, the control `dataset' of joint responses generated via this procedure corresponds with a population of neurons that fire independently, but have heterogeneous firing rates that match the original data.

\subsection{B.2 Homogeneous Firing Rates Shuffled Control}
To generate this control, we first computed the average mean rate from the data, $\big \langle r_i \big \rangle_{i=1}^N$, averaged over the entire population of $N$ ganglion cells, where 
$r_i \equiv \left( \sum_{t=1}^T \sigma_i(t) \right)/T$.  
We then simulated an independent population of matched size $N$ in which each neuron $i$ was assigned to have its firing rate $r_i = \rho \equiv \big \langle r_i \big \rangle_{i=1}^N$. 
Thus, the control `dataset' of joint responses generated via this procedure corresponds to a population of neurons that fire independently, and moreover have homogeneous firing rates.

\subsection{B.3 Results for the Control Analyses}
\begin{figure*}[!ht]
\centering
\includegraphics[scale=0.5]{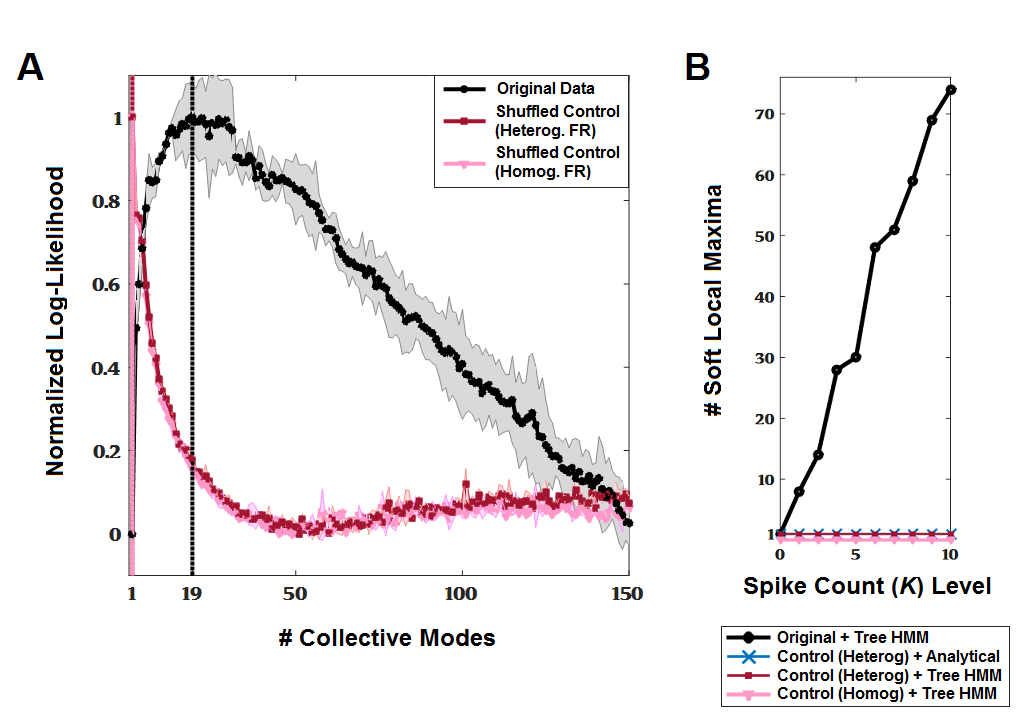}
\caption{Control results. 
(A) Normalized cross-validated log-likelihood (CV-LL) as a function of the number of HMM collective modes for: the original dataset of 152 ganglion cell responses to Movie $\#2$ (black), the heterogeneous firing rate shuffled control (red), and the homogeneous firing rate shuffled control (pink).  
Shown in bold is the mean normalized CV-LL over all $n$ cross-validation folds (in practice taken to be $n = 2$). Shaded error bars denote SEM over cross-validation folds. Each colored dashed line indicates the
optimal latent dimensionality for the corresponding dataset (see Key for colors). 
(B) Shown is the number of unique $K$-soft local maxima identified via opposite sign neuron pair relaxation, as a function of spike count $K$. Results are shown for four cases: the original dataset with the Tree HMM chosen as the underlying probability model (black circles); the heterogeneous firing rate shuffled control with the analytical independent model as the underlying probability model (blue crosses); the heterogeneous firing rate shuffled control with the fit Tree HMM as the
underlying probability model (red squares); and the homogeneous firing rate shuffled
control with the Tree HMM as the underlying probability model (pink triangles).
}
\label{fig11}
\end{figure*}

As seen in Fig \ref{fig11}A, there was a categorical difference between the cross-validated log-likelihood (CV-LL) curves obtained when the HMM was fit to the original data, versus when the HMM was fit to the control ``data".  
Whereas the CV-LL curve for the original data exhibited a well-defined peak at 19 collective modes, when fit to each control dataset, the CV-LL curve instead achieved its maximum at only one collective mode.

The soft local maxima results were likewise substantially different for the original data versus the controls (Fig \ref{fig11}B).  In particular, there was a monotonic increase in the proliferation of soft local maxima at all spike count levels for the original data, with 74 unique soft local maxima identified at spike count level $K=10$.   
In contrast, the modeled probability landscapes for both controls exhibited either zero or one unique soft local maximum across all spike count levels.  

Specifically, for the first control, in which the ganglion cell firing rates in the original data were preserved and thus heterogeneous across the population, one unique soft local maximum was present at each spike count level.  
As expected analytically for an independent neuron population with heterogeneous firing rates, we confirmed that the single soft local maximum identified at each spike count level $K$ corresponded with the $K$ ganglion cells that had the highest firing rates.   
This scenario corresponds with a ganglion cell population that is organized into one, single neuronal community. 
In this case, the ridge union graph corresponds to the complete graph (i.e. it has all-to-all connectivity between all nodes). 
Thus, in the case of a population of heterogeneous neurons that fire independently, there is no community structure.  

For the second control, in which each neuron was moreover assigned the same firing rate, no soft local maxima were found for any of the tested spike count levels.  
This is consistent with the analytical result and expectation for an independent population of homogeneous neurons, as the symmetry in the homogeneous case implies that the probability landscape depends only on $K$ (and thus all joint responses within the same spike count level have identical probability).  

In summary, results from the control analyses support that the soft local maxima results - and by extension the ridge and neuronal community results - are non-trivially dependent on the empirical correlation structure of the measured neural activity. 

\onecolumngrid
\section{Appendix C: Supplementary Figures}

\begin{figure}[!h]
\centering
\includegraphics[width = 1\textwidth]{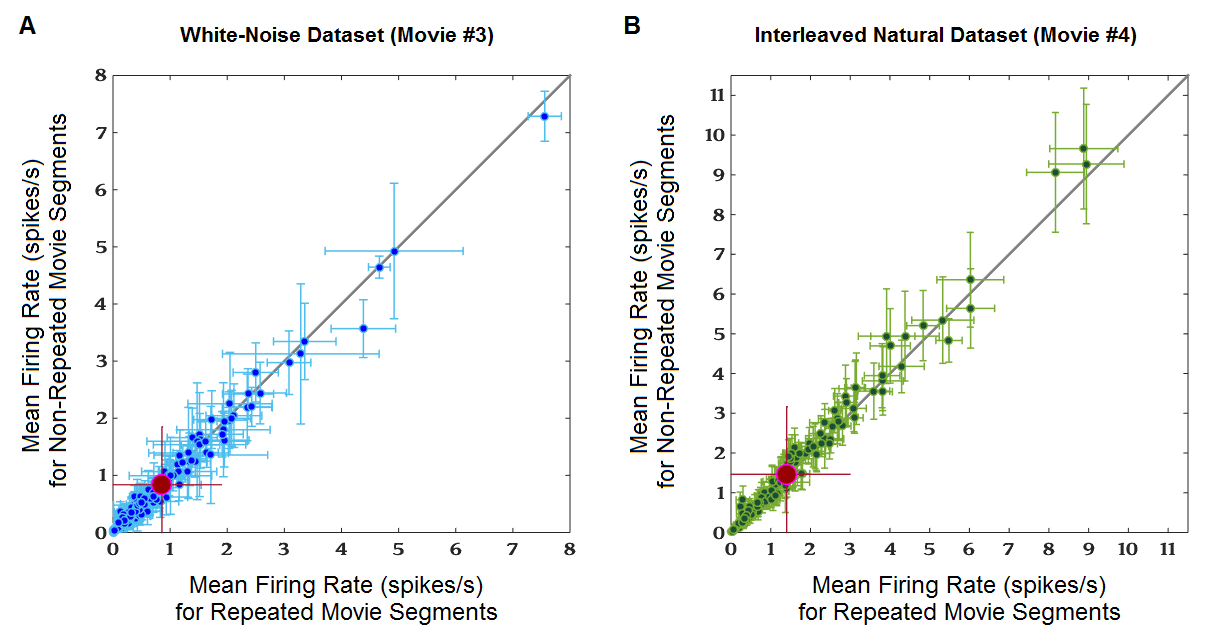}
\caption{Scatter plot of mean firing rates for the raw data. 
(A) Results for the dataset of $N=155$ ganglion cells responding to the interleaved white-noise checkerboard stimulus ensemble (Movie $\#3$).  Each blue dot represents the raw firing rate (spikes/s) for one ganglion cell. Shown is each neuron's mean firing rate during the \emph{non-repeated} movie segments, averaged over all unique movie segments comprising the original Movie $\#3$ ($y$-axis), versus its mean firing rate during the \emph{repeated} movie segments ($x$-axis).  Error bars denote one standard deviation over movie segments. Gray line denotes the line of unity.  The enlarged red dot denotes the mean firing rate across the entire population of $155$ ganglion cells; error bars denote one standard deviation over ganglion cells. 
(B) Results for the dataset of $N=170$ ganglion cells responding to the interleaved natural movie stimulus ensemble (Movie $\#4$).  The format is the same as in Panel (A). 
}
\label{fig:S1}
\end{figure}

$\:$\\
$\:$\\
$\:$\\
$\:$\\
$\:$\\
$\:$\\

\begin{figure}[!h]
\centering
\includegraphics[width = 1.05\textwidth]{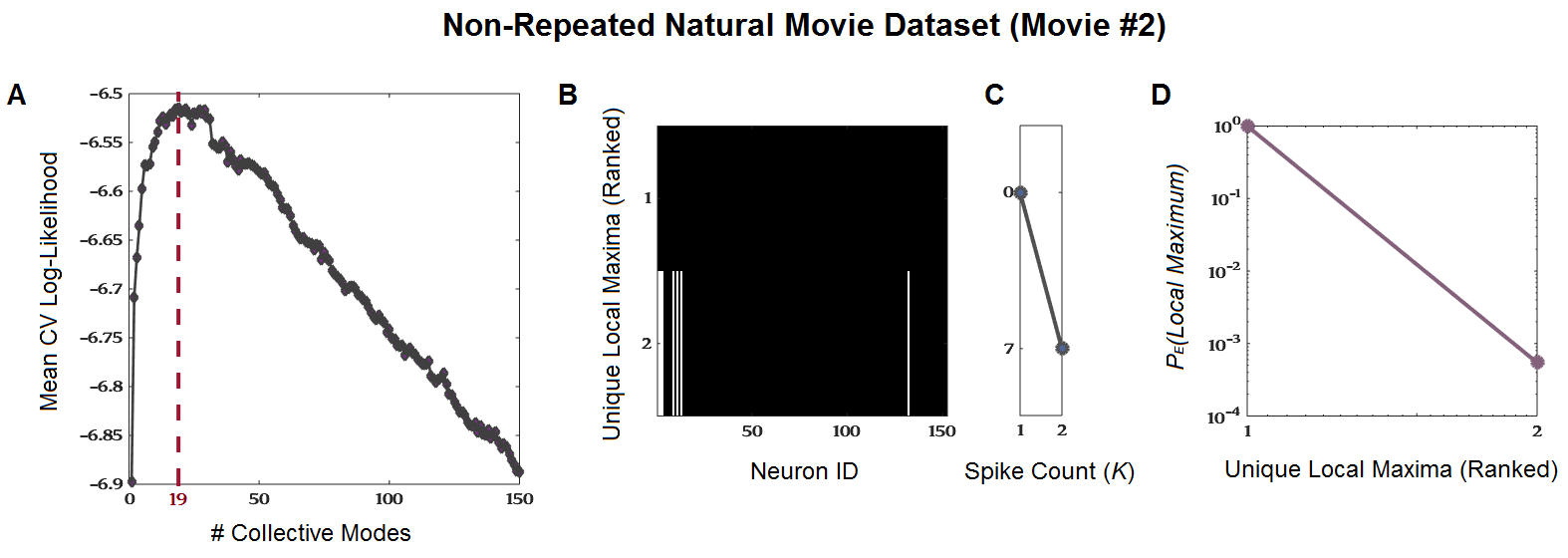}
\caption{Local maxima results for the dataset of 152 ganglion cells responding to Movie $\#2$. 
(A) Cross-validated log-likelihood (CV-LL) averaged over the two cross-validation folds ($y$-axis), as a function of the number of Tree HMM latent states (collective modes) ($x$-axis). 
Red dashed line denotes the optimal latent dimensionality, which corresponds with the peak of the CV-LL curve. 
(B) Each row represents the binary representation of an identified unique soft local maximum. 
(C) The spike count of each associated soft local maximum shown in panel (B). 
(D) Plot of the proportion (denoted by $P_E$; note the log scale) of the 90,001 population responses observed in the data which were mapped via the single spin flip ascent algorithm to the corresponding soft local maximum, indicated on the $x$-axis (ranked).}
\label{fig:S2}
\end{figure}

$\:$\\
$\:$\\
$\:$\\
$\:$\\
$\:$\\
$\:$\\
$\:$\\
$\:$\\
$\:$\\
$\:$\\
$\:$\\

\begin{figure}[!h] 
\centering
\includegraphics[width = 1\textwidth]{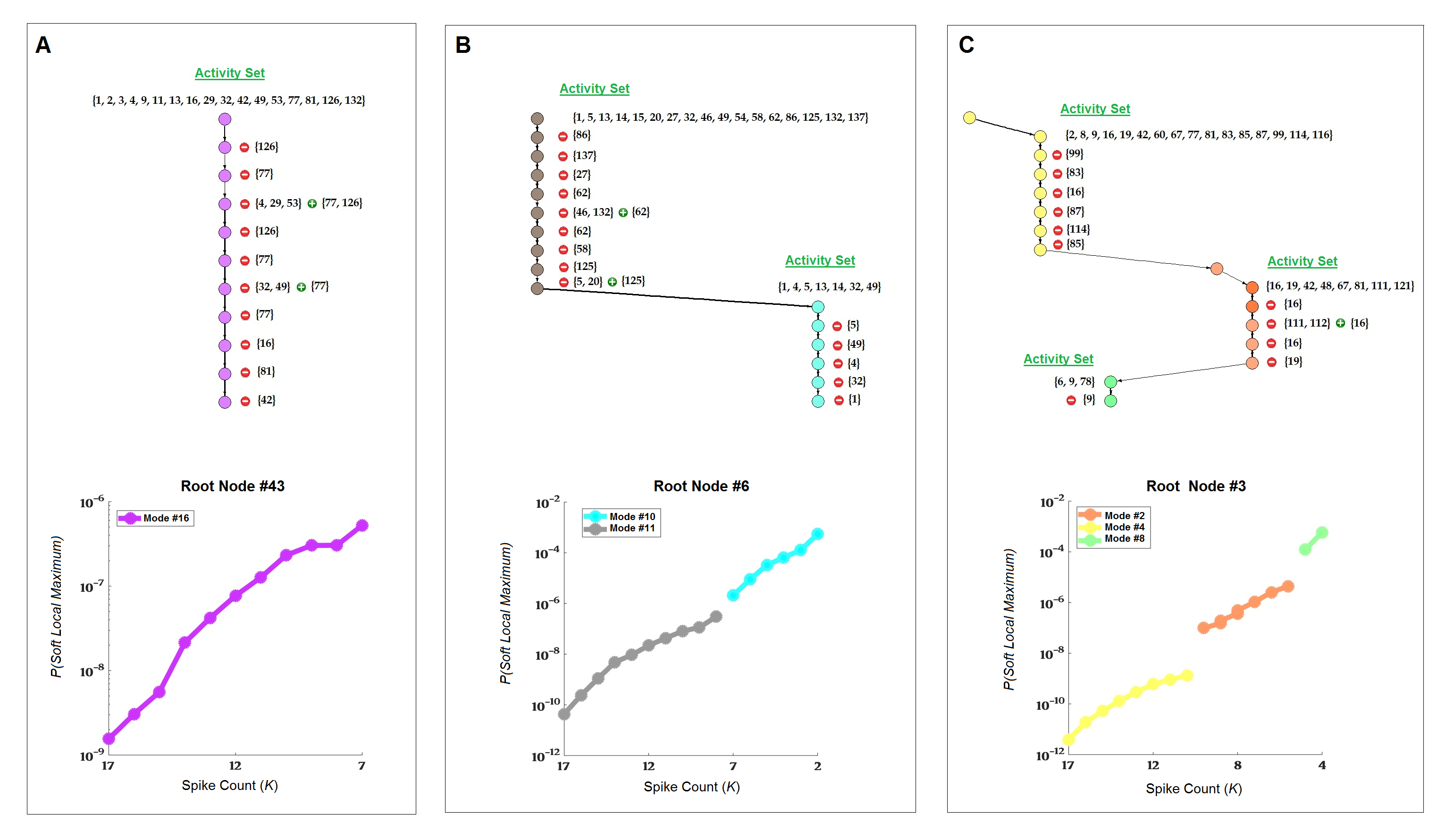}
\caption{ Examples of output rooted digraphs using d-reachable edges. 
(A) Output rooted digraph obtained via our ridge search algorithm, described in the main text, when we took the unique ($K=17$)-soft local maximum with rank $43$ as the input root node.  Notation is the same as in Figs 8 and 9. 
The rooted weighted digraph obtained when the $(K=17)$-soft local maximum with rank $6$ and $3$ was taken as the input root node is shown in (B) and (C), respectively. 
}
\label{fig:S8}
\end{figure}

\newpage


\end{document}